\documentclass[a4paper,fleqn,usenatbib]{mnras}

\usepackage{newtxtext,newtxmath}

\usepackage[T1]{fontenc}
\usepackage{ae,aecompl}

\usepackage{hyperref}
\usepackage[small,bf]{caption}
\usepackage{natbib}
\usepackage{amsmath}
\usepackage{wasysym}
\usepackage{rotating}
\usepackage{multirow}

\newcommand{\fig}[1]{Fig.~\ref{fig:#1}}
\newcommand{\Table}[1]{Table~\ref{tab:#1}}
\newcommand{\se}[1]{\S\ref{sec:#1}}

\newcommand{\hi}{H\footnotesize{I}\relax}

\title{An Extensive Photometric Catalog of CALIFA Galaxies}

\author[C. Gilhuly \& S. Courteau]{
Colleen Gilhuly,$^{1,2}$\thanks{E-mail: gilhuly@astro.utoronto.ca (CG)}
and St\'ephane Courteau$^{1}$
\\
$^{1}$Department of Physics, Engineering Physics \& Astronomy, Queen's University, Kingston, ON K7L 3N6, Canada\\
$^{2}$Department of Astronomy \& Astrophysics, University of Toronto, Toronto, M5S 3H4, Canada
}

\pubyear{2018}

\begin{document}
\label{firstpage}
\pagerange{\pageref{firstpage}--\pageref{lastpage}}
\maketitle

\begin{abstract}
We present an extensive compendium of photometrically-determined structural properties for all CALIFA 
galaxies in the Third Data Release (DR3). We exploit Sloan Digital Sky Survey (SDSS) images in order 
to extract one-dimensional (1D) $gri$ surface brightness profiles for all CALIFA DR3 galaxies.  We also 
derive a variety of non-parametric quantities and parametric models fitted to 1D $i$-band profiles.  
The galaxy images are decomposed using the 2D bulge-disc decomposition programs \textsc{imfit} and 
 \textsc{galfit}.  The relative performance and merit of our 1D and 2D modelling approaches 
are assessed. 
Where possible, we compare and augment our photometry with existing measurements from the literature. 
Close agreement is generally found with the studies of \cite{walcher14} and \cite{mendezabreu17}, 
though some significant differences  exist.
Various structural metrics are also highlighted on account of their tight dispersion against an independent 
variable, such as the circular velocity.
\end{abstract}




\section{Introduction}
\label{sec:intro}

Extragalactic astronomy rests on a very rich foundation of catalogs of
galaxy parameters (e.g. CGCG, UGC, RSA, PGC, RC3, SDSS/NYU-VAGC)  
\citep[to cite only a few]{zwicky68, nilson73, sandage81, paturel89, deVaucouleurs91, nyu_value}.
These catalogs have fueled a most impressive array of investigations ranging from galaxy classification,
bulge and disc parameterizations, identification of underlying components, scaling relations, evolution
of structural properties with time, and the coupling with spectroscopically-determined parameters.

The most comprehensive understanding of galaxies is achieved when structural, dynamical, and chemical
information are combined. The latter two rely mostly on spectroscopic measurements. However, unlike their
photometric counterparts, large spectroscopic catalogs of galaxies lack in numbers, given the relatively
longer single-object exposures (compared to multi-target broadband photometry) and the multiple
long-slit or integral field unit (IFU) pointings required per galaxy.  Fortunately, the advent of integral
field spectroscopy (IFS) surveys such as ATLAS3D, CALIFA, SAMI, and MaNGA 
\citep{cappellari11,sanchez12,walcher14,bryant15,bundy15}, have enabled spatially-resolved
spectroscopic observations for many thousands of galaxies, and this trend is only growing.
Needless to say, the combination of statistically-significant photometric and spectroscopic samples
greatly minimizes selection biases and enables mature astrophysical investigations
(e.g. such as multi-layered galaxy scaling relations). 

The Calar Alto Legacy Integral Field spectroscopy Area (CALIFA) survey \citep{sanchez12,walcher14}
exemplifies this new trend.  It uses a size-selected sample spanning a variety of environments in the
local universe (out to z $\sim$ 0.03).  Size selection allows the most efficient use of the IFU, good
coverage of targets, and ensures a uniform statistical sampling of spatially resolved properties such
as age, metallicity, and surface mass density. CALIFA's sample is large enough to achieve some statistical 
significance yet still small enough to allow detailed single wide-field IFU mapping per individual galaxy. 
Thus, CALIFA plays a crucial role in providing a local spectroscopic baseline for future extragalactic surveys 
at higher redshift. CALIFA datacubes offer better spatial resolution, spatial coverage and signal-to-noise 
than other surveys such as MANGA and SAMI, thus highest overall S/N. On the other hand, CALIFA's sample
size is smaller than that of MaNGA or SAMI, its spectral resolution is equivalent to or lower, and its
spectral coverage is smaller than MaNGA's \citep{sanchez12, walcher14, bryant15, bundy15}. Although
smaller in sample size than typical imaging surveys, CALIFA's spatial coverage and resolution ensure its 
ability to probe spatially resolved structures such as HII regions and spiral arms. CALIFA was deemed
complete after its third and final data release (DR3) in April 2016 \citep{sanchez16b}, comprising 667 galaxies.  

The philosophy of enhancing spectroscopic studies with photometric parameters guides our current effort.
Indeed, this work endeavors to provide the most complete compendium of photometric properties for
all galaxies covered in the third CALIFA data release. We also contrast our photometry and value-added 
parameters with those from existing databases. 

In \se{data}, we present the sample and the SDSS data used to extract surface brightness profiles. 
In \se{profiles}, the process of extracting those profiles as well as structural parameters from those profiles
is explained. \se{models} introduces our 1D and 2D parametric models, which are then contrasted against
each other in \se{compare_1D_2D}. Finally, we extensively
compare our final catalog with existing CALIFA photometry in \se{compare} and conclude with likely
applications of our new database in \se{discuss}. 

\section{Data and Photometry}
\label{sec:data}

Our study covers the full third CALIFA data release, with its Main Sample (MS) of 542 galaxies 
\citep[see][]{walcher14} and another 125 galaxies from the Extension Sample \citep[ES;][]{sanchez16b}.
$10' \times 10'$ cut-out
images of each CALIFA DR3 galaxy were extracted from the Sloan Digital Sky Survey (SDSS) Data
Release 10 (DR10) online mosaic interface \citep{sdssdr8,sdssdr10} using SWarp \citep{SWarp}. The large
image area ensures that the sky background can be well characterized for all CALIFA galaxies.
Images in all five SDSS bands were retrieved, but our analysis focuses on $gri$ photometry.
In particular, the $i$-band is chosen as the preferred band for the extraction of structural parameters
since it probes the longest wavelengths and thus offers the best views of the underlying mass distribution
of galaxies whilst minimizing the impact of dust in the stellar atmospheres and our own Earth's glowing atmosphere. 

Source Extractor \citep{bertin96} was used to detect the galaxy, stars, and background sources
in the $i$-band image. The raw Source Extractor galaxy masks were slightly expanded and smoothed, and any
disjoint regions were removed from the masks. The detected stars and background source masks were
also slightly expanded and smoothed, and used to mask the images during profile extraction and 
2D galaxy modelling (see \se{profiles} and \se{2D_models}).

In the presence of very bright stars, close and/or overwhelmingly bright neighbour galaxies,
or ongoing mergers, source detection and masking are weakened.  Seventeen galaxies were
affected by these issues and have been flagged as poorly masked. While these galaxies have
been processed as per normal, and can be found in our public database, they are routinely
excluded from our analyses.

SDSS DR10 imaging is already sky-subtracted; issues regarding over-subtraction around
large galaxies have previously been acknowledged and addressed, though some of the
lost flux in SDSS photometry may be attributed to deblending rather than sky errors \citep{blanton11}.
We compute the residual sky level, avoiding any biases in the SDSS DR10 sky subtraction pipeline.

\section{Surface brightness profile extraction}
\label{sec:profiles}

For the production of 
azimuthally-averaged surface brightness profiles, we take advantage of XVISTA\footnote{
http://ganymede.nmsu.edu/holtz/xvista/}, a suite of image analysis
programs for astronomical applications. A detailed description of this software
and data modelling procedures used in this study is  presented in \cite{courteau96};
complementary descriptions of profile extraction procedures are also found 
in \cite{mcdonald11} and \cite{hall12}.

The process of profile extraction begins with the initialization of variables describing
image parameters, determination of the sky level (measured in boxes located near the image corners), 
and calculation of the photometric centre of the galaxy. The first isophotal solution is
then calculated over the visible extent of the galaxy. The position angle and ellipticity of 
outer contours are poorly constrained due to the low surface brightness
at those radii.  Therefore, a reliable mid-to-outer contour is adopted to define the position angle
and ellipticity of the outer contours. Typically there is little variation in contour parameters in the stable
regions where the best contours are selected, so the precision of contour selection introduces minimal
uncertainty to the outer profile.

The user may smooth the isophotal solution to avoid contour crossings and excursions
due to non-axisymmetric structures, such as bars and spiral structure.  
Despite the subjective nature of this procedure, \cite{hall12} verified that independent operators
obtain similar surface brightness profiles.  We also show in \se{compare} that parameters obtained
from our profile extraction method are very similar to those obtained from growth curves with fixed
position angle and ellipticity \citep{walcher14}. These comparisons demonstrate that any bias introduced
from best isophote selection and subsequent smoothing is minimal.

Once the $i$-band isophotal solution is finalized, final corrections and transformations
are applied to obtain a surface brightness profile in AB magnitude units. Our final $i$-band
isophotal contours are then imposed on to images in other bands and their surface brightness
profiles are also produced.  The common isophotal solution for all bands ensures uniform
profile comparisons and physically meaningful colour profiles.

Surface brightness profiles that extend far into the sky noise-dominated regime are typically
truncated when the SB error exceeds 1.5 mag arcsec$^{-2}$.  The $gri$ surface
brightness profiles for a selection of CALIFA galaxies are presented in Appendix~\ref{sec:gri_profiles}.

\subsection{Parameter extraction}
\label{sec:phot}

Isophotal radii and enclosed total magnitudes are determined for several isophotal levels in each band. 
As in \cite{hall12}, we use the $i$-band $\mu = 23.5$ mag arcsec$^{-2}$ level to characterize a galaxy's outer regions. 
Intrinsic SB errors rise rapidly beyond this isophotal radius \citep{courteau96}.
The resulting isophotal parameters, $R_{23.5}$ and $m_{23.5}$, require no assumptions
or parameterizations of the galaxy light distribution.  In addition to the isophotal magnitude,
the total magnitude is calculated.

Effective or half-light parameters $R_e$ and $\mu_e = \mu(R_e)$ are defined
by the radius which encloses half of the total light of the galaxy.  The latter is
computed from the galaxy's growth curve integrated from the surface brightness profile.
Similarly to $R_{23.5}$ and $m_{23.5}$, these parameters are model-independent.

The radii enclosing 20 and 80 per cent of the total light are also determined 
and used to calculate the concentration index $C_{28} =5 \log (R_{80}/R_{20})$.
For a pure exponential disc galaxy, $C_{28} = 2.8$; for a de Vaucouleurs profile, $C_{28} = 4$.

Stellar masses are calculated according to mass-to-light versus colour relations (MLCRs)
tabulated by \cite{roediger15}.  They advocate for the use of multiple colours to
better constrain $\Upsilon_*$; this advice is echoed by \cite{zhang17}.
We select the four MLCRS based on \cite{bruzual03} stellar population models
with the highest Pearson correlation coefficients ($g-r,g$; $g-r,r$; $g-i,g$; $g-i,r$)
and average the resulting stellar masses. We then calculate $\Upsilon_*(i)$ 
using the average stellar masses and our $i$-band total magnitudes. We produce these
stellar masses to extend our basic photometry and to facilitate further comparison
with \cite{walcher14}. For an extended discussion of MLCRs, the many assumptions
and choices associated with their determination, and a comparison of published
transformations, see \cite{zhang17}.

\cite{roediger15} found that stellar masses have systematic
uncertainty of $\sim$0.3 dex due to assumptions in the stellar population modelling
\citep{Conroy13,Courteau14}. An additional uncertainty of 0.06-0.07 dex results
from reliance on integrated quantities rather than radially or spatially resolved maps. Therefore,
we assume 0.31 dex or 71 per cent systematic uncertainty for stellar masses. This dominates
over our estimated random measurement uncertainty of 24 per cent (\se{error}). 

In addition to the structural measurements derived from surface brightness profiles,
the Gini coefficient $G$ and the moment of light $M_{20}$ are calculated \citep{gini1912,lotz04}.
Both are non-parametric descriptions of the distribution of intensity amongst galaxy pixels,
akin to but distinct from light concentration. Unlike the non-parametric effective or isophotal
quantities, neither $G$ nor $M_{20}$ require any assumptions regarding the symmetry of the galaxy.
They are thus a most generic and non-parametric structure descriptors, and are most fittingly
described as \emph{symmetry-independent}.

Studies of distant galaxies relying on $G$ and $M_{20}$ highlight their value when resolution
is low and circular symmetry or a clear centre may be absent \citep{abraham03, lotz04}. 
In the low redshift regime probed by the mostly undisturbed CALIFA galaxies, $G$ and $M_{20}$ 
may offer no advantage over $C_{28}$.  However, we have opted to present them here for their legacy
value in the context of comparative studies against higher redshift galaxies.

While $G$ was invented as a measure of economic equality \citep{gini1912},
astronomers have embraced it to describe the relative equity of the distribution of light, particularly 
for classification purposes \citep[and references therein]{abraham03,lotz04,lisker08}
\footnote{In a population (or galaxy image) with each individual (pixel) possessing the same wealth 
(flux), $G = 0$. If all the wealth (or flux) is concentrated in one individual (pixel), $G = 1$. 
Geometrically, $G$ is proportional to the area contained between the cumulative distribution
function of the population ranked by increasing wealth and the distribution in a population
with complete equality (a straight line).}.
Unlike many other measurements of galactic structure, $G$ does not require any
photometric centre. $G$ can be calculated using the following formula \citep{glasser62}:

\begin{equation}
\label{eq:gini}
G = \frac{1}{\bar{X}n(n-1)}\sum_{i=1}^n (2i - n - 1)X_i \textrm{   } (n > 2).
\end{equation}

Here, the $X_i$ values are the individual pixel fluxes in order of increasing flux
and $\bar{X}$ is the mean pixel flux.  $X_i$ is replaced with its absolute value to improve 
robustness against noise \citep{lotz04}. 

$M_{20}$ is the second-order moment of light of the brightest 20 per cent of the galaxy's total flux,
normalized by the total second-order moment of light $M_{tot}$ \citep{lotz04}. $M_{tot}$ is
defined as follows:
\begin{equation}
M_{tot} = \sum^n_i M_i = \sum^n_i f_i[(x_i - x_c)^2 + (y_i - y_c)^2],
\end{equation}

\noindent
where $f_i$ is the flux of the $i$th pixel, and $x_c$ and $y_c$ define the galaxy's centre,
and are determined such that $M_{tot}$ is minimized. Furthermore,
\begin{equation}
\label{eq:M20}
M_{20} = \log \Bigg( \frac{\sum_i M_i}{M_{tot}}\Bigg)\textrm{, while }\sum_i f_i< 0.2f_{tot}.
\end{equation}

Both $G$ and $M_{20}$ require the pixels of a galaxy to be explicitly designated. The choice
of an edge for extended sources such as galaxies is a challenging problem in image analysis,
and biases our resulting measurements. Fortunately, the previously described galaxy masks
determined by Source Extractor provide a convenient way to automate the process
of galaxy segmentation and minimize human intervention and thus bias. 

\subsection{Parameter corrections}
\label{sec:corr}

The parameters described above have been corrected according to procedures described in
\cite{courteau07} and \cite{hall12}. One major deviation is the lack of a K-correction in this work. 
Since all of the CALIFA Main Sample galaxies (542 out of the 667 CALIFA DR3 galaxies) lie at $z < 0.03$,
 the K-correction is $\sim$ 0.01 mag and can be safely ignored. This K-correction for the most distant 
galaxy in the CALIFA Main Sample would be $\sim$0.05 mag; still small enough to justify neglecting the 
correction for the whole sample; other corrections are far larger and uncertain.  While some of the 
CALIFA Extension Sample galaxies lie at redshifts of $\sim 0.1$, we still neglect their K-correction. 

The total magnitudes in each band are corrected for Galactic extinction using the 
\cite{schlafly11} maps based on SDSS stellar spectra.  The Galactic extinction for each galaxy was
retrieved using the NASA/IPAC Extragalactic Database (NED)\footnote{http://ned.ipac.caltech.edu/}. 
Our adopted prescription for correcting the internal extinction of spiral galaxies
originates from \cite{tully98}, with the following form:
\begin{equation}
A_{\lambda} = \gamma_\lambda \log(a/b),
\end{equation}

\noindent
where $a/b$ is the axial ratio of the galaxy and the  wavelength-dependent $\gamma_\lambda$ are
\begin{align}
\gamma_g = 1.51 + 2.46(\log W - 2.5), \\
\gamma_r = 1.25 + 2.04(\log W - 2.5),\\
\gamma_i = 1.00 + 1.71(\log W - 2.5).
\end{align}

Deprojected \hi~ line widths from \cite{springob05} were nominally used for $W$. When
unavailable, stellar and gaseous velocity dispersions \citep[][Gilhuly et al., in prep]{sanchez16}
empirically corrected to approximate these line widths were used.

Both the Galactic and internal extinction corrections amount to $\sim$0.1 mag, increasing the observed
galaxy brightness by a few tenths of a magnitude. Absolute magnitudes are calculated from the corrected
apparent magnitudes according to
\begin{equation}
M = m - 5 \log (D_L/\textrm{kpc}) + 25.
\end{equation}

A Python cosmology calculator \citep{wright06} is used to calculate $D_L$ from the NED redshift, 
taking $H_0 = 70 \textrm{ km s}^{-1} \textrm{ Mpc}^{-1}$. Following \cite{hall12},
each of the parameters required for correction in Equations 4 and 8 (as well as Galactic
extinction) have an assumed 10-15 per cent uncertainty. This yields a typical systematic uncertainty
of 0.3 mag for corrected $M_{23.5}$ and total magnitudes. 

Surface brightnesses are corrected for Galactic and internal extinction as well as cosmological
dimming: 
\begin{equation}
\mu_e'= \mu_e + 0.5 \log (a/b) - A_G - 2.5 \log(1+z)^3. 
\end{equation}

Radius corrections, as reported by \cite{courteau07}, do not result in a reduction of scatter
for scaling relations and we elect to use uncorrected radii given the uncertainties in the corrections. 
We follow their lead and do not attempt any radius corrections.

\subsection{Parameter uncertainties}
\label{sec:error}

In order to assign typical uncertainties to the structural parameters presented in this work,
we posit that sky background variations present the most significant source of random error.
Sky uncertainty in the sky-subtracted DR10 images is assumed to be $\pm 5$ times the residual sky, 
and we inspect 65 galaxies ($\sim 10$ per cent of CALIFA DR3) where this assumed uncertainty
leads to moderate changes in the outer surface brightness profile.  
The sky error sample is representative of the diversity of morphology in CALIFA,
as gauged by the range of $C_{28}$. 

With their nearly flat, asymptotically
declining, outer surface brightness profiles, early-type galaxies are most greatly
impacted by sky variations.  Since our error estimates apply to all galaxy types,
these are likely over-estimated for late-type galaxies.

The surface brightness profiles with over- and under-estimated sky backgrounds, at levels corresponding
to $\pm 5$ times the residual sky, are processed exactly the same way as the original profiles
to arrive at the same set of measurements described in \se{phot}.  The relative difference from
the original measurements is tabulated and 3-sigma outliers/tails are rejected accordingly.
The median difference of each parameter is recalculated using the trimmed set
and is taken as the characteristic relative error, shown in \Table{phot_err}.  
The 8 per cent uncertainty on total $i$-band magnitudes due to sky background errors translate
into an 11 per cent uncertainty on colours (e.g. $g-i$ and $g-r$), a 24 per cent uncertainty on stellar masses,
and a 25 per cent uncertainty on $M_*/L$.

For $G$ and $M_{20}$, the determination of the ``edge" of the galaxy pixel region presents a major 
source of systematic error. A number of galaxy mask schemes could be determined: using the raw
Source Extractor mask, slightly smaller or larger mask dilation during smoothing, or various isophotal
levels (24.5 mag arcsec$^{-2}$, 25 mag arcsec$^{-2}$, or 25.5 mag arcsec$^{-2}$). The largest relative
difference between our fiducial measurements and those calculated using alternate galaxy masks
is tabulated and the median of the central distribution is found as above. Error estimates
are shown in \Table{phot_err}.  The uncertainty for $M_{20}$ and $G$ are 44 and 14 per cent, respectively,
indicating that $G$ is less sensitive to the exact galaxy edge definition.

These uncertainties are significantly larger than those quoted by \cite{lotz04} for $M_{20}$ (10 per cent relative error; 
by comparing results from different photometric sources) and both \cite{abraham03} and \cite{lotz04} for
$G$ (4 per cent while using bootstrap resampling and comparison of measurements from different photometric sources, respectively). 
This indicates that photometric variation entails far smaller errors than the definition of a galaxy's edge, especially for $M_{20}$.

\begin{table}
\begin{centering}
\begin{tabular}{c c c c c c c c }
\hline \hline
$M_{23.5}$ & $M_{tot}$ &  $\mu_e$  & $R_{23.5}$ & $R_e$ & $C_{28}$ & $G$ & $M_{20}$\\ \hline
5\% & 8\% & 12\% & 6\% & 9\% & 3\% & 14\% & 44\% \\

\hline \hline
\end{tabular}
\caption{Estimated uncertainties of non-parametric photometric quantities.}
\label{tab:phot_err}
\end{centering}
\end{table}

\subsection{Total and extrapolated quantities}

We have presented in \se{phot} the isophotal and total parameters extracted from surface brightness profiles
in our photometric catalog.   Recall that total parameters are typically measured at a radius where the surface
brightness error is $\sim$1.5 mag arcsec$^{-2}$ in each band.  It is also common practice to extrapolate
a surface brightness profile to infinity using a suitable S\'ersic function.  For spiral galaxies, 
\cite{courteau96} and \cite{hall12} use an exponential disc extrapolation to estimate total magnitudes.  Model 
magnitudes may also be extracted by integrating idealized profiles to infinity. Extrapolating the surface 
brightness profile to infinity naturally impacts integrated quantities such as the total magnitude and colours,
as well as effective quantities like $R_e$ and $\mu_e$, or concentrations.  

Profile extrapolations may increase the effective radius, $R_e$, by 10-15 per cent and concentrations, $C_{28}$, 
by as much as 20 per cent. Total $i$-band magnitudes will typically change (brighten) by 0.1 mag or less and 
yield bluer colours by the same amount.  Stellar masses calculated with color mass-to-light relations (hereafter, 
CMLRs) will also change accordingly.  The practice of extrapolating a surface brightness profile to infinity may 
seem arbitrary; the method of identifying the end of a profile is somewhat subjective as well.  

In light of the relatively small differences between total and extrapolated parameters, 
we focus on the former for our catalog and scientific investigations. 

\subsection{Photometry table}
\label{sec:phot_table}

\begin{table*}
\begin{center}
{\small
\begin{tabular}{ccccccccccccc}
\hline \hline
\multirow{3}{*}{Name} &  \multirow{2}{*}{$M_{23.5}$}  & \multirow{2}{*}{$M_{i}$} & \multirow{2}{*}{$g-r$} & 
\multirow{2}{*}{$g-i$} & $R_{23.5}$ & $R_e$ &  \multirow{2}{*}{$\mu_e$} & \multirow{2}{*}{$C_{28}$} & 
\multirow{2}{*}{$G$} & \multirow{2}{*}{$M_{20}$} & log$M_\ast$ & \multirow{2}{*}{$\Upsilon_*(i)$} \\
&  &  &  &  & (kpc) & (kpc) &  &  &  &  & ($M_\odot$) &\\
& $\pm 5\%$ & $\pm 8\%$ & $\pm 11\%$ & $\pm 11\%$ & $\pm 6\%$ & $\pm 9\%$ & $\pm 12\%$& $\pm 3\%$ & $\pm 14\%$ & $\pm 44\%$ &  $\pm 24\%$  & $\pm 25\%$ \\
\hline \hline
IC5376 & -21.95 & -22.06 & 0.59 & 0.86 & 17.07 & 5.35 & 21.06 & 5.65 & 0.69 & -2.87 & 10.71 & 1.06 \\  
UGC00005 & -22.76 & -22.85 & 0.54 & 0.84 & 20.69 & 8.05 & 20.74 & 2.61 & 0.52 & -1.72 & 10.96 & 0.92 \\ 
NGC7819 & -21.35 & -21.56 & 0.48 & 0.76 & 14.99 & 9.17 & 22.02 & 3.78 & 0.46 & -2.38 & 10.37 & 0.76 \\  
UGC00029 & -22.67 & -22.98 & 0.82 & 1.24 & 22.36 & 10.11 & 21.72 & 5.12 & 0.57 & -2.52 & 11.41 & 2.28 \\ 
IC1528 & -21.66 & -21.74 & 0.40 & 0.60 & 14.64 & 6.23 & 21.09 & 3.11 & 0.52 & -1.83 & 10.31 & 0.57 \\ \hline 
NGC7824 & -22.87 & -22.92 & 0.76 & 1.13 & 20.52 & 5.26 & 20.68 & 5.13 & 0.60 & -2.77 & 11.31 & 1.90 \\  
UGC00036 & -22.62 & -22.73 & 0.68 & 1.02 & 19.63 & 5.01 & 20.31 & 4.32 & 0.67 & -2.41 & 11.11 & 1.45 \\  
NGC0001 & -21.93 & -22.11 & 0.65 & 1.02 & 13.38 & 3.66 & 20.54 & 5.25 & 0.65 & -2.64 & 10.82 & 1.32 \\ 
NGC0014 & -18.69 & -19.03 & 0.48 & 0.57 & 4.20 & 2.36 & 22.28 & 3.42 & 0.36 & -1.64 & 9.32 & 0.71 \\  
NGC0023 & -22.78 & -22.89 & 0.66 & 1.00 & 17.14 & 6.03 & 19.88 & 4.95 & 0.68 & -2.50 & 11.15 & 1.35 \\ \hline 
NGC0036 & -22.90 & -23.05 & 0.59 & 0.87 & 24.10 & 9.21 & 21.10 & 4.19 & 0.53 & -2.25 & 11.10 & 1.05 \\ 
UGC00139 & -20.69 & -20.94 & 0.34 & 0.50 & 10.45 & 5.67 & 21.82 & 3.60 & 0.54 & -1.98 & 9.90 & 0.46 \\  
UGC00148 & -21.39 & -21.48 & 0.42 & 0.57 & 13.83 & 6.48 & 20.80 & 3.19 & 0.59 & -1.68 & 10.22 & 0.59 \\  
MCG-02-02-030 & -21.52 & -21.59 & 0.59 & 0.88 & 12.55 & 4.81 & 20.40 & 3.54 & 0.60 & -2.12 & 10.52 & 1.06 \\  
UGC00312NOTES01 & -20.51 & -20.63 & 0.52 & 0.73 & 6.78 & 2.10 & 20.26 & 4.21 & 0.67 & -2.24 & 10.03 & 0.84 \\ \hline 
ESO539-G014 & -21.79 & -21.89 & 0.22 & 0.24 & 20.64 & 9.01 & 21.69 & 2.89 & 0.52 & -1.91 & 10.10 & 0.30 \\  
MCG-02-02-040 & -20.81 & -21.02 & 0.53 & 0.78 & 11.95 & 6.02 & 21.29 & 2.71 & 0.53 & -1.53 & 10.21 & 0.87 \\  
UGC00335NED02 & -21.47 & -21.85 & 0.83 & 1.27 & 13.93 & 7.23 & 21.94 & 5.41 & 0.57 & -2.62 & 10.98 & 2.42 \\ 
NGC0155 & -22.68 & -22.90 & 0.84 & 1.28 & 21.74 & 8.20 & 21.20 & 4.77 & 0.58 & -2.46 & 11.42 & 2.49 \\  
ESO540-G003 & -21.05 & -21.16 & 0.41 & 0.65 & 10.16 & 3.39 & 20.65 & 3.70 & 0.61 & -2.10 & 10.09 & 0.59 \\ \hline 
UGC00355 & -21.57 & -21.62 & 0.70 & 1.09 & 15.66 & 5.35 & 20.72 & 3.54 & 0.52 & -1.97 & 10.71 & 1.57 \\  
NGC0160 & -22.97 & -23.05 & 0.71 & 1.06 & 25.44 & 7.83 & 21.43 & 4.83 & 0.51 & -2.68 & 11.27 & 1.56 \\ 
NGC0165 & -21.94 & -22.11 & 0.60 & 0.85 & 15.22 & 7.72 & 21.66 & 3.04 & 0.46 & -2.09 & 10.74 & 1.09 \\  
NGC0169 & -22.75 & -22.78 & 0.83 & 1.22 & 23.50 & 6.55 & 20.60 & 4.42 & 0.59 & -2.11 & 11.35 & 2.37 \\  
NGC0171 & -22.24 & -22.29 & 0.60 & 0.93 & 15.00 & 7.93 & 20.92 & 2.83 & 0.46 & -2.08 & 10.83 & 1.13 \\ \hline 
NGC0177 & -21.98 & -22.02 & 0.42 & 0.59 & 15.70 & 3.24 & 21.13 & 5.53 & 0.61 & -2.82 & 10.45 & 0.60 \\ 
NGC0180 & -22.65 & -22.80 & 0.58 & 0.93 & 22.38 & 11.25 & 21.49 & 2.58 & 0.42 & -1.95 & 11.01 & 1.06 \\  
NGC0192 & -22.27 & -22.38 & 0.74 & 1.09 & 22.22 & 6.86 & 20.34 & 4.97 & 0.65 & -2.63 & 11.05 & 1.75 \\  
NGC0216 & -19.67 & -19.77 & 0.33 & 0.45 & 5.93 & 2.40 & 20.84 & 2.72 & 0.57 & -1.65 & 9.41 & 0.44 \\  
NGC0214 & -22.57 & -22.61 & 0.60 & 0.91 & 16.83 & 5.47 & 20.39 & 3.41 & 0.57 & -2.14 & 10.94 & 1.09 \\ \hline 
NGC0217 & -22.61 & -22.68 & 0.72 & 1.06 & 24.72 & 7.17 & 20.30 & 4.60 & 0.60 & -2.41 & 11.15 & 1.63 \\  
MCG-02-02-086 & -24.26 & -24.58 & 0.92 & 1.37 & 46.70 & 25.10 & 21.54 & 3.58 & 0.43 & -1.73 & 12.20 & 3.19 \\  
NGC0237 & -21.45 & -21.56 & 0.54 & 0.79 & 11.10 & 4.37 & 20.57 & 3.39 & 0.58 & -2.00 & 10.44 & 0.90 \\ 
NGC0234 & -22.23 & -22.32 & 0.57 & 0.91 & 15.76 & 6.63 & 20.88 & 2.43 & 0.47 & -1.80 & 10.80 & 1.02 \\  
MCG-02-03-015 & -22.62 & -22.69 & 0.53 & 0.75 & 19.93 & 5.39 & 20.44 & 5.29 & 0.63 & -2.75 & 10.87 & 0.85 \\ \hline 
NGC0257 & -22.66 & -22.74 & 0.56 & 0.83 & 19.61 & 7.89 & 20.81 & 3.36 & 0.53 & -2.17 & 10.94 & 0.95 \\  
IC1602 & -23.89 & -24.18 & 0.94 & 1.39 & 39.45 & 17.24 & 21.47 & 4.64 & 0.54 & -2.30 & 12.07 & 3.42 \\  
NGC0309 & -22.89 & -23.02 & 0.53 & 0.77 & 25.02 & 12.76 & 21.34 & 2.62 & 0.37 & -1.64 & 11.00 & 0.86 \\ 
NGC0364 & -21.95 & -22.02 & 0.84 & 1.27 & 14.46 & 4.17 & 20.79 & 5.07 & 0.60 & -2.71 & 11.06 & 2.45 \\ 
NGC0426 & -21.93 & -22.12 & 0.81 & 1.29 & 13.83 & 3.80 & 20.58 & 5.92 & 0.67 & -2.79 & 11.07 & 2.28 \\ \hline 
NGC0429 & -22.15 & -22.20 & 0.65 & 0.92 & 17.42 & 2.46 & 19.45 & 5.05 & 0.76 & -2.83 & 10.85 & 1.28 \\  
IC1652 & -21.51 & -21.58 & 0.74 & 1.13 & 16.16 & 4.00 & 19.87 & 5.09 & 0.69 & -2.60 & 10.74 & 1.79 \\  
NGC0447 & -22.58 & -22.82 & 0.81 & 1.16 & 22.90 & 11.93 & 21.67 & 4.94 & 0.53 & -2.55 & 11.32 & 2.15 \\  
NGC0444 & -20.99 & -21.13 & 0.36 & 0.57 & 16.91 & 8.74 & 21.95 & 3.10 & 0.45 & -1.84 & 10.01 & 0.50 \\  
UGC00809 & -20.55 & -20.68 & 0.38 & 0.54 & 12.01 & 5.93 & 21.52 & 3.13 & 0.47 & -1.59 & 9.85 & 0.52 \\ \hline 
UGC00841 & -21.08 & -21.17 & 0.44 & 0.69 & 15.32 & 6.63 & 21.48 & 2.74 & 0.50 & -1.73 & 10.14 & 0.65 \\  
NGC0472 & -21.77 & -22.01 & 0.85 & 1.21 & 12.65 & 3.72 & 20.89 & 6.03 & 0.67 & -2.82 & 11.06 & 2.48 \\ 
CGCG536-030 & -20.48 & -20.55 & 0.35 & 0.54 & 8.42 & 3.73 & 21.19 & 2.63 & 0.50 & -1.56 & 9.76 & 0.48 \\ \hline
\hline
\end{tabular} }

\caption{Catalog of corrected photometry described in \protect\se{phot}. All errors
are estimated using sky variation, except $G$ and $M_{20}$, whose errors are estimated
based on the uncertainty in defining a galaxy's edge. Systematic errors are not
included here. Poorly masked galaxies are excluded from this table, but are
included in the full table with a quality flag.  
The full version of this table is available online as supplementary material, as well
as at: 
\url{https://www.physics.queensu.ca/Astro/people/Stephane_Courteau/gilhuly2017/index.html}.} 

\label{tab:phot}

\end{center}
\end{table*}

Having discussed the processing of surface brightness profiles and galaxy pixel maps,
we now present a summary of our table of photometric parameters in \Table{phot}. 
The full catalog (available as supplementary online material and at 
\url{https://www.physics.queensu.ca/Astro/people/Stephane_Courteau/gilhuly2017/index.html}) 
integrates our photometry, parametric modelling (see \se{models} and \Table{models}),
right ascension, declination, redshift, position angle, ellipticity, quality flags, 
and relevant ancillary data such as Hubble types \citep{walcher14} and velocities \citep{springob05}.  
This catalog will serve as an authoritative reference for CALIFA galaxies, in support 
of the exploitation of its rich spectroscopic observations. The condensed
version presented here includes the following information: 

\medskip

Column 1: Galaxy name.

\smallskip

Column 2: $M_{23.5}$, the total absolute $i$-band magnitude contained within the 23.5 mag arcsec$^{-2}$
isophotal level.

\smallskip

Column 3: $M_{i}$, the total absolute $i$-band magnitude integrated over the entire surface brightness profile.

\smallskip

Column 4: $g-r$, the colour calculated from the total magnitudes (Col. 3) in the $g$ and $r$ bands.

\smallskip

Column 5: $g-i$, the colour calculated from the total magnitudes (Col. 3) in the $g$ and $i$ bands.

\smallskip

Column 6: $R_{23.5}$, the semi-major axis length of the $i$-band 23.5 mag arcsec$^{-2}$ isophote in kpc.

\smallskip

Column 7: $R_e$, the semi-major axis length of the $i$-band isophote enclosing half of the galaxy's total light.

\smallskip

Column 8: $\mu_e$, the effective $i$-band surface brightness at $R_e$.

\smallskip

Column 9: $C_{28}$, the concentration index relating the radii enclosing 20 and 80 per cent of the galaxy's total light.

\smallskip

Column 10: $G$, the Gini coefficient, a symmetry-independent concentration measurement
for the equality of light distribution amongst galaxy pixels in the $i$-band image \citep{abraham03,lotz04,lisker08}.

\smallskip

Column 11: $M_{20}$, a symmetry-independent concentration measurement for the second-order
moment of light of the brightest galaxy pixels that together contribute 20 per cent of the galaxy's total light \citep{lotz04}.

\smallskip

Column 12: log $M_\ast$/$M_\odot$, the stellar mass of the galaxy determined from an average of
four MLCRs \citep{roediger15}.  The full catalog also contains log $M_{\ast, 23.5}$/$M_\odot$, 
the stellar mass calculated as above but limited to $R < R_{23.5}$.

\smallskip

Column 13: $\Upsilon_*(i)$, the total $i$-band stellar mass to light ratio. The full catalog also contains $\Upsilon_{*,23.5}(i)$, 
the stellar mass to light ratio within $R < R_{23.5}$.

\section{1D and 2D modelling}
\label{sec:models}

We have presented in \se{profiles} a non-parametric investigation of galaxy light profiles, extracting 
quantities such as effective radii and surface brightness, concentration, Gini coefficients, moments of 
the light distribution, galaxy colour, and stellar masses.  In order to achieve a most complete structural 
description of all CALIFA galaxies, we now turn to parametric descriptions of the galaxy light.
  
Galaxy one-dimensional (1D) surface brightness profiles and two-dimensional (2D) images may 
be fitted with idealized parametric functions in order to provide a common framework for the structural 
analysis of galaxies of all types.  Such decomposition methods have been devised and applied for 1D 
profiles \citep{kent85,baggett98,macarthur03,mcdonald11} and 2D images in multiple bands
\citep{byun95,dejong96,desouza04,galfit,simard11,lacknergunn12,erwin15,mendezabreu17,robotham17}. 

1D surface brightness profiles provide the variation of intensity, position angle and ellipticity
with radius; therefore, parametric models based on surface brightness profiles naturally allow for
variation in position angle and ellipticity according to the degree of smoothing enforced during
profile extraction. Most 2D photometric modelling programs do not include tilted rings and
require a fixed position angle and ellipticity for each structual component. This fundamental
difference may impact the final fitted parameters, justifying the comparison of 1D and 2D models
in a homogeneous way. 

A promising related approach also involves full spectral decompositions of spectral data cubes. Indeed, 
the photometric decompositions of SDSS images for CALIFA galaxies presented below deserve extensive 
comparison with those obtained from full spectral decompositions of CALIFA data cubes \citep{tabor17}.  
The latter is however beyond the scope of this paper.

We now examine a suite of decomposition models for our SDSS galaxy 1D light profiles and 2D images, 
based on the overall understanding that galaxies are typically composed of disc-like and spheroid-like
components modeled respectively by exponential and S\'ersic profiles.  For a quick review, the exponential
function in magnitude units is defined as:

\begin{equation}
\label{eq:exp}
\mu(R) = \mu_0 + 1.086 \Bigg(\frac{R}{h}\Bigg),
\end{equation}

where $\mu_0$ is the central surface brightness and $h$ is the exponential scale length.  

For spheroid-like systems, the S\'ersic function \citep{sersic63} is expressed as

\begin{equation}
\label{eq:src}
\mu(R) = \mu_e + 1.086 b_n \Bigg[\Bigg(\frac{r}{R_e}\Bigg)^{\frac{1}{n}} - 1 \Bigg],
\end{equation}

where $R_e$ is the half-light radius and  $b_n$ is the S\'ersic coefficient. We adopt the calculation of $b_n$
by \cite{macarthur03}. During our 2D modelling, $n$ is left as a free parameter despite concerns that 
it may be poorly constrained \citep{macarthur03,mcdonald09ursa,simard11}. The stability of our solutions below
is assessed in \se{bd_ratios} in order to determine the impact of this choice. 

The functions that we adopt to model our galaxy light profiles include (i) single-component models 
(Exponential and S\'ersic), and (ii) two-component bulge-disc models (Exponential + Exponential 
[or ``Double Exponential''] and S\'ersic + Exponential). The structural models are fitted to both 
the $i$-band surface brightness profiles (1D) and to the original SDSS $i$-band images (2D).  
Of the four fitted models, the preferred 1D and 2D models were selected according to criteria to be 
explained below.

Simple models are favoured over more complex multi-component models in the interest of automation
and the assessment of the success of large scale bulge-disc decomposition studies. An arbitrarily large number
of fitted components may yield a perfect data-model match, though potentially at the expense of physical meaning.
While it is possible to study smaller samples of galaxies in greater detail, closely supervising fits and injecting
components to describe rings, spiral arms, lenses, bars, and other features, such studies become interactively
taxing with samples of hundreds or thousands of galaxies.  Still, given model degeneracies and the vagaries
of B/D decompositions, we still caution against automated decompositions of large galaxy samples and
recommend close monitoring. The goal of our our modelling effort is to seek a balance between optimal
galaxy parameterizations and minimal user interaction.

All 1D and 2D models were assigned a basic quality flag, indicating the representativeness 
of the model to the galaxy's light distribution. These quality flags were assigned through brief
visual inspection of the model components plotted against the galaxy's surface brightness profiles.
Some leniency was employed when assessing 2D models. The quality flag levels are:

\begin{itemize}
\item{0: Model failed to converge normally. Do not use these models.}
\item{1: Model has converged but either fits very little of the galaxy, has one of two components
at vanishingly low surface brightness (degenerate with a single-component model), or has abused/misused/swapped
model components relative to the intended bulge-disc paradigm. These models may be usable with caution.}
\item{2: Model describes only inner regions of galaxy well, describes most or all of the galaxy but
poorly, or involves slight misuse/abuse of model components
relative to the intended bulge-disc paradigm.}
\item{3: Model is well-behaved and fairly descriptive of galaxy over most or all of its visual extent.}
\end{itemize}

Note that these flags do not constitute an assessment of the validity of a model, and
there is considerable subjectivity and overlap in these quality levels. In general, we recommend
that users consider models with quality flags of 2 or 3, but some applications may be more
selective or more permissive with regard to model quality.

From the pool of acceptable models (quality flags or 2 or 3), the model with the lowest reduced $\chi^2$ is
noted as our ``preferred'' model. When a galaxy has no models with sufficiently high quality flags, no
model is identified as preferred. This is done separately for the 1D, \textsc{imfit}, and \textsc{galfit} models, resulting in
one preferred model per modelling code per galaxy.  This selection technique is very simplistic; users may devise their
own criteria for identifying the most appropriate model for a given galaxy as all our models and their
parameters have been preserved.

The final parameters for all four models in 1D and 2D are included in the photometry catalog
along with the quantities described in \se{phot}. Corrections to surface brightnesses are done
following the procedure in \se{corr}. 

\subsection{1D model fitting}
\label{sec:1D_models}

We make use of the nonlinear least-squares optimization implemented in the scipy.optimize
Python module \citep{scipy} to fit 1D models to surface brightness profiles.
The initial parameters for the effective radii and surface brightnesses are based on previous non-parametric estimates.  
A grid search about the initial parameters is used due to the sensitivity of optimization algorithms to the
starting position in parameter space.  We include 1D convolution with a  radially symmetric point spread function 
(PSF) in our user-defined functions fitted to the surface brightness profiles, in order to account for atmospheric 
blurring. The surface brightness profile is mirrored about $R=0$ to yield a symmetric profile.

A double Gaussian function is used for the PSF, following the SDSS pipelines 
convention\footnote{https://www.sdss3.org/dr10/algorithms/magnitudes.php}. 
The following equation governs the PSF:

\begin{equation}\label{eq:PSF}
f(x) =  C_1 e^{-\frac{1}{2}(\frac{x-x_0}{\sigma_1})^2} + C_2 e^{-\frac{1}{2}(\frac{x-x_0}{\sigma_2})^2},
\end{equation}

\noindent
where $C_1$ and $C_2$ are normalization constants, $x_0$ is the centre and peak of the PSF (common to both Gaussian functions),
and $\sigma_1$ and $\sigma_2$ are the standard deviations of each Gaussian function. $C_1$ and $\sigma_1$ correspond to the
peaked core of the PSF while $C_2$ and $\sigma_2$ describe the wings of the PSF.
{Up to 20 bright but unsaturated stars per image were fitted with this PSF model, and
the median values of $C_1/C_2$, $\sigma_1$, and $\sigma_2$ from individual PSF fits
were taken as the best PSF parameters for the image. Typical values for these parameters
are 8, 0.40$''$, and 1.0$''$ respectively. We find significant PSF variations across SDSS
images (up to $\sim 30$ per cent for $\sigma_{1,2}$) and confirm that individually-fitted PSFs
are required to ensure the quality of 1D and 2D photometric models, particularly
for bulge parameters.

The 1D fitting script checks for undesirable solutions. Any negative, exceptionally
large, or NaN parameters cause the solution to be rejected. For two-component models, the bulge component
is also prevented from being brighter than the disc component at the maximum radius of the input profile.
Solutions that pass these checks have their reduced $\chi^2$ evaluated. 
If the reduced $\chi^2$ is lower than that of the previous best solution, the new
solution is saved as the best solution and the grid search continues. 

For the S\'ersic + Exponential model, the S\'ersic index $n$ of the bulge component
was kept fixed during fitting runs. Values from 0.2 - 7 were tested, roughly following \cite{macarthur03}. 
This imposes some restrictions on the shape of the bulge, which is generally poorly constrained.

For models with bulge components, an inner reduced $\chi^2$ is calculated to ensure that the bulge is well
fitted \citep{macarthur03}. As the region of bulge dominance is typically small compared to the full radial
extent of the profile, it is possible that a solution that fits an outer feature (such as a spiral arm or break)
would have a lower global reduced $\chi^2$ than a solution appropriately modelling the bulge. The default
inner region is defined as $1 R_e$. This value has lower and upper limits of 20 pixels (corresponding to
approximately 3 kpc for CALIFA galaxies) and half of the total profile 
length, respectively. Instead of selecting the best model as the model with the lowest global reduced 
$\chi^2$, the models are ranked by increasing global and inner reduced $\chi^2$. The model that has the 
lowest sum of squared ranks is selected as the best model; a compromise between
a good global fit and a good inner fit rather than fitting one region much better at the expense of the other.

\subsection{2D model fitting}
\label{sec:2D_models} 

The image modelling code \textsc{imfit} was first singled out for this investigation.
\textsc{imfit} distinguishes itself from other publicly available codes by its open source, object-oriented design
that makes defining custom image components straightforward. 

Previously determined background source masks (described in \se{data}) are used. Parameters 
defining image uncertainties such as gain, read noise, and original sky level are described using a global
estimate rather than one per image due to the lack of pertinent information in the mosaicked image headers. 
Information regarding the original sky level was not preserved in the SDSS DR10 cut-out images, making
a return to the corresponding  non sky-subtracted DR7 images necessary to estimate the typical original sky
($\sim 1$ nMgy; note that nMgy is a relative unit of flux, defined such that 
$m = 22.5 \textrm{ mag} - 2.5 \log_{10}(f/\textrm{nMgy})$). The gain was also determined from DR7 image headers ($\sim 1000 e^-/$nMgy). 
The read noise ($\sim 6.5 e^-$) was obtained from NASA-Sloan Atlas\footnote{http://www.nsatlas.org/} (NSA)
cut-out images, a catalog of nearby galaxies constructed from SDSS imaging with updated background subtraction \citep{blanton11}.
In the light of our experience, we recommend the use of DR7 imaging for 2D modelling.

The imaging parameters can be obtained
from previous data releases since the images are produced from the same observations but processed
in different ways; it is merely inconvenient that these parameters are not all included in the cut-out image headers.
Slight variations of these image parameters did not affect the final fitted parameters during test runs of \textsc{imfit};
the use of a single estimate for all galaxies is therefore not expected to significantly affect the resulting
models. The parameter found to have the largest galaxy-to-galaxy variation
is the original sky, varying over a range of approximately 0.5 nMgy.  

However, the normalization of fit statistics such as the reduced $\chi^2$ is affected by the final
values for the estimated gain, read noise, and original sky. Furthermore, the noise characteristics
of the mosaic images are different from that of a single frame, and we cannot account for this difference.
Therefore, the fit statistics of our \textsc{imfit} models cannot be directly compared to those of our 1D and \textsc{galfit}
models. Fortunately, this has minimal impact on the fit parameters themselves or for selecting the
\textsc{imfit} model that best describes a galaxy. Varying the imaging
parameters involved in sigma image generation naturally changes the fit statistic normalization but leaves
the best fitted parameters unchanged. This indicates that the sigma image generated by \textsc{imfit}
is sensible.  

The best-fitting parameters from the corresponding 1D model were used to initialize the \textsc{imfit} runs.
This shielded us against poor initial parameters leading to unphysical minima in parameter space. 
The 1D fits benefited from a grid search which would be too time consuming for 2D modelling.
PSF convolution and a fixed flat sky component was included, the latter previously measured in two strips
350 pixels wide at the top and bottom of each (masked) image.

To supplement and further validate the 2D \textsc{imfit} models, we also tested the popular 2D decomposition
software \textsc{galfit} \citep{galfit}.  \textsc{galfit} has been widely exploited to parameterize galaxies
\citep{galfit1,galfit2,galfit3,galfit4,galfit5,galfit6,galfit7,galfit8}
which bolsters our comparisons with our 1D and \textsc{imfit} models.
The above discussion about image parameters, masking, fit initialization, and parameter constraints
holds for \textsc{galfit} models as well, with a few exceptions. \textsc{galfit} internally estimates the 
sky/background error  of an image in regions devoid of bright pixels. This approach obviates the
need for our estimated read noise (which is otherwise not representative of the noise in the mosaics),
and the normalization of the reduced $\chi^2$ is approximately correct.

\subsection{Model parameter uncertainties}
\label{sec:model_err}

As in \se{error}, we assume that the uncertainty in the sky level dominates the
random error in our fitted model parameters. Only galaxies with a given
preferred model are used to determine the impact of sky uncertainty on that model.
For 1D models, the surface brightness profiles extracted with purposefully over- 
and under-estimated sky backgrounds are refitted to obtain
the range of likely structural parameters.  For 2D models, the constant sky component is increased
and held fixed while the previous best fit parameters are used to initialize the next modelling iteration.
Once again, the median relative difference is used to define the uncertainty estimates.
These error estimates are shown in Table~\ref{tab:model_err}.
The errors are comparable to the non-parametric measurement errors in \se{error} but the \textsc{imfit} errors
are often smaller. This is due to the previously-noted reduced sensitivity
to low surface brightness features of 2D models (as compared to 1D models),
leading to a reduced sensitivity to variations in the sky level.

\begin{table}
\begin{centering}
\begin{tabular}{c  c c c c c c c c c }
\hline \hline
Model & $\mu_{B}$ & $R_{B}$ & $n_B$ & $\mu_D$  & $R_{D}$ &  $n_D$ \\ \hline \hline
\multirow{3}{*}{Exp}  & -- & -- & -- & 7\% & 5\% & -- \\
 & -- & -- & -- & 1\%  & 1\% & -- \\
 & -- & -- & -- & 7\%  & 5\% & -- \\
\\
\multirow{3}{*}{Ser} &  -- & -- & -- & 26\% & 13\% & 18\% \\
 &  -- & -- & -- & 36\% & 24\% & 13\% \\
 &  -- & -- & -- & 51\% & 23\% & 23\% \\
\\
\multirow{3}{*}{Exp+Exp}  & 9\% & 8\% & -- & 14\% & 8\% & -- \\
 & 1\%  & 1\%  & -- & 2\%  & 3\%  & -- \\
 & 6\%  & 6\%  & -- & 15\%  & 10\%  & -- \\
\\
\multirow{3}{*}{Ser+Exp}  & 20\% & 15\% & 17\% & 20\% & 12\% & -- \\
 & 7\% & 5\% & 4\% & 4\%  & 4\% & -- \\
 & 27\% & 13\% & 13\% & 24\%  & 15\% & -- \\
\hline \hline
\end{tabular}
\caption{Estimated uncertainties of the single-component and bulge-disc models. The 1D model 
errors are specified first, with the \textsc{imfit} and \textsc{galfit} model errors in the following rows.
For single-component models, parameters are recorded in the disc columns
(notated by a subscript ``D'').}
\label{tab:model_err}
\end{centering}
\end{table}

\subsection{Preferred models catalog}

\begin{table*}
\begin{center}
{\small
\begin{tabular}{ccccccccccc}
\hline\hline
\multirow{2}{*}{Name} &  Model & $\mu_B$ & $R_B$ & \multirow{2}{*}{$n_B$} & $\mu_D$ & $R_{1,D}$ & $R_{2,D}$ & $R_{brk,D}$ & \multirow{2}{*}{$\alpha_D$} & \multirow{2}{*}{$n_D$} \\
& (1D, IF, GF) & (mag arcsec$^{-2}$) & (kpc) &  & (mag arcsec$^{-2}$) & (kpc) & (kpc) & (kpc) & & \\
\hline\hline
\multirow{3}{*}{IC5376} & Exp+Exp & 17.22 & 0.74 & -- & 19.94 & 5.34 & -- & -- & -- & -- \\ 
 & Src+Exp & 19.46 & 1.61 & 4.26 & 20.40 & 5.91 & -- & -- & -- & -- \\ 
 & Src+Exp & 22.40 & 7.95 & 9.19 & 20.60 & 5.20 & -- & -- & -- & -- \\ \hline 
\multirow{3}{*}{UGC00005} & Exp & -- & -- & -- & 19.01 & 5.20 & -- & -- & -- & -- \\ 
 & Src & -- & -- & -- & 20.92 & 9.05 & -- & -- & -- & 1.00 \\ 
 & Exp+Exp & 17.48 & 0.33 & -- & 19.18 & 5.64 & -- & -- & -- & -- \\ \hline 
\multirow{3}{*}{NGC7819} & Exp+Exp & 17.56 & 0.44 & -- & 20.31 & 5.43 & -- & -- & -- & -- \\ 
 & Src+Exp & 19.43 & 0.86 & 1.26 & 20.61 & 5.89 & -- & -- & -- & -- \\ 
 & Src+Exp & 19.43 & 0.87 & 1.28 & 20.73 & 5.92 & -- & -- & -- & -- \\ \hline 
\multirow{3}{*}{UGC00029} & Src & -- & -- & -- & 23.05 & 20.26 & -- & -- & -- & 5.10 \\ 
 & Src+Exp & 20.37 & 3.62 & 3.12 & 21.36 & 12.65 & -- & -- & -- & -- \\ 
 & Src+Exp & 20.17 & 3.26 & 2.87 & 21.54 & 12.93 & -- & -- & -- & -- \\ \hline 
\multirow{3}{*}{IC1528} & Exp & -- & -- & -- & 19.27 & 3.61 & -- & -- & -- & -- \\ 
 & Src+Exp & 20.36 & 0.61 & 1.12 & 19.49 & 4.04 & -- & -- & -- & -- \\ 
 & Exp+Exp & 18.43 & 0.36 & -- & 19.50 & 4.07 & -- & -- & -- & -- \\ \hline 
\multirow{3}{*}{NGC7824} & Exp & -- & -- & -- & 18.85 & 4.75 & -- & -- & -- & -- \\ 
 & Src+Exp & 18.74 & 1.59 & 2.93 & 20.18 & 5.81 & -- & -- & -- & -- \\ 
 & Src+Exp & 18.54 & 1.46 & 2.71 & 20.24 & 6.74 & -- & -- & -- & -- \\ \hline 
\multirow{3}{*}{UGC00036} & Src & -- & -- & -- & 21.14 & 7.30 & -- & -- & -- & 2.38 \\ 
 & Exp+Exp & 16.19 & 0.29 & -- & 18.59 & 2.83 & -- & -- & -- & -- \\ 
 & Exp+Exp & 16.22 & 0.32 & -- & 18.69 & 3.25 & -- & -- & -- & -- \\ \hline 
\multirow{3}{*}{NGC0001} & Src & -- & -- & -- & 21.32 & 5.49 & -- & -- & -- & 4.69 \\ 
 & Src & -- & -- & -- & 21.23 & 5.13 & -- & -- & -- & 4.98 \\ 
 & Src & -- & -- & -- & 21.38 & 5.54 & -- & -- & -- & 5.19 \\ \hline 
\multirow{3}{*}{NGC0014} & Exp+Exp & 19.71 & 0.77 & -- & 22.04 & 2.81 & -- & -- & -- & -- \\ 
 & Exp+Exp & 20.29 & 0.56 & -- & 20.35 & 1.36 & -- & -- & -- & -- \\ 
 & Exp+Exp & 19.74 & 0.79 & -- & 21.65 & 1.94 & -- & -- & -- & -- \\ \hline 
\multirow{3}{*}{NGC0023} & Exp+Exp & 17.18 & 2.12 & -- & 20.81 & 7.29 & -- & -- & -- & -- \\ 
 & Src+Exp & 18.53 & 1.60 & 3.77 & 19.14 & 5.04 & -- & -- & -- & -- \\ 
 & Exp+Exp & 14.95 & 0.47 & -- & 18.57 & 4.10 & -- & -- & -- & -- \\ \hline 
\multirow{3}{*}{NGC0036} & Exp+Exp & 17.30 & 0.86 & -- & 20.01 & 8.12 & -- & -- & -- & -- \\ 
 & Src+Exp & 21.44 & 3.81 & 4.61 & 19.99 & 6.79 & -- & -- & -- & -- \\ 
 & Src+Exp & 21.71 & 5.47 & 4.67 & 20.31 & 8.17 & -- & -- & -- & -- \\ \hline 
\multirow{3}{*}{UGC00139} & Src & -- & -- & -- & 22.31 & 6.85 & -- & -- & -- & 2.06 \\ 
 & Src & -- & -- & -- & 21.85 & 5.42 & -- & -- & -- & 1.72 \\ 
 & Src & -- & -- & -- & 21.87 & 5.50 & -- & -- & -- & 1.74 \\ \hline 
\multirow{3}{*}{UGC00148} & Exp & -- & -- & -- & 19.09 & 3.51 & -- & -- & -- & -- \\ 
 & Src & -- & -- & -- & 20.74 & 6.15 & -- & -- & -- & 0.70 \\ 
 & Src+Exp & 21.53 & 2.71 & 0.32 & 19.17 & 3.48 & -- & -- & -- & -- \\ \hline 
\multirow{3}{*}{MCG-02-02-030} & Exp & -- & -- & -- & 18.81 & 2.97 & -- & -- & -- & -- \\ 
 & Exp+Exp & 17.65 & 0.30 & -- & 18.94 & 2.85 & -- & -- & -- & -- \\ 
 & Exp+Exp & 17.48 & 0.25 & -- & 18.83 & 2.98 & -- & -- & -- & -- \\ \hline 
\multirow{3}{*}{UGC00312NOTES01} & Src & -- & -- & -- & 20.55 & 2.33 & -- & -- & -- & 2.43 \\ 
 & Src+Exp & 18.91 & 0.63 & 0.87 & 19.40 & 1.89 & -- & -- & -- & -- \\ 
 & Exp+Exp & 16.75 & 0.34 & -- & 18.66 & 1.39 & -- & -- & -- & -- \\ \hline 
\multirow{3}{*}{ESO539-G014} & Exp & -- & -- & -- & 19.93 & 5.79 & -- & -- & -- & -- \\ 
 & Exp+Exp & 18.62 & 0.75 & -- & 19.88 & 5.79 & -- & -- & -- & -- \\ 
 & Src & -- & -- & -- & 21.87 & 9.96 & -- & -- & -- & 1.41 \\ \hline 

\hline
\end{tabular} }
\caption{Preferred 1D and 2D \textsc{imfit} (IF) and \textsc{galfit} (GF) model parameters. 
Poorly masked galaxies were excluded from this table,
but are included in the full table with a quality flag.
The full version of this table, and the parameters for all models, are available online as
supplementary material and at 
\url{https://www.physics.queensu.ca/Astro/people/Stephane_Courteau/gilhuly2017/index.html}.
}
\label{tab:models}
\end{center}
\end{table*}

Our preferred models and their parameters for each modelling code are tabulated in \Table{models}. Our full
online catalog includes all four models for each modelling code and each galaxy as well as their reduced
$\chi^2$. The models are collected alongside the photometry presented in \Table{phot}, along with
the additional information presented in \se{phot_table}. The columns of \Table{models} are as follows:

\medskip

Column 1: Galaxy name.

\smallskip

Column 2: Preferred models identified from the suite of 1D, \textsc{imfit}, and \textsc{galfit} models.

\smallskip  

Column 3: $\mu_B$, the characteristic surface brightness of the bulge component where relevant: $\mu_0$ for
exponential bulges and $\mu_e$ for S\'ersic bulges.

\smallskip

Column 4: $R_B$, the scale radius of the bulge component in kpc where relevant: $h_b$ for exponential bulges
and $R_{e,b}$ for S\'ersic bulges.

\smallskip

Column 5: $n_B$, the S\'ersic index of the bulge component, where relevant.

\smallskip

Column 6: $\mu_D$, the characteristic surface brightness of a single-component model or the disc component of a
bulge-disc model: $\mu_0$ for exponential discs and $mu_e$ for single S\'ersic models.

\smallskip

Column 7: $R_{D}$, the scale radius of a single-component model or the disc component of a bulge-disc model in kpc:
$h$ for exponential discs, $R_e$ for single S\'ersic models.

\smallskip

Column 8: $n_D$, the S\'ersic index of a single S\'ersic model, where relevant.

\section{Comparison of 1D and 2D decompositions}
\label{sec:compare_1D_2D}

\begin{figure}
\begin{centering}
\includegraphics[width=0.49\textwidth]{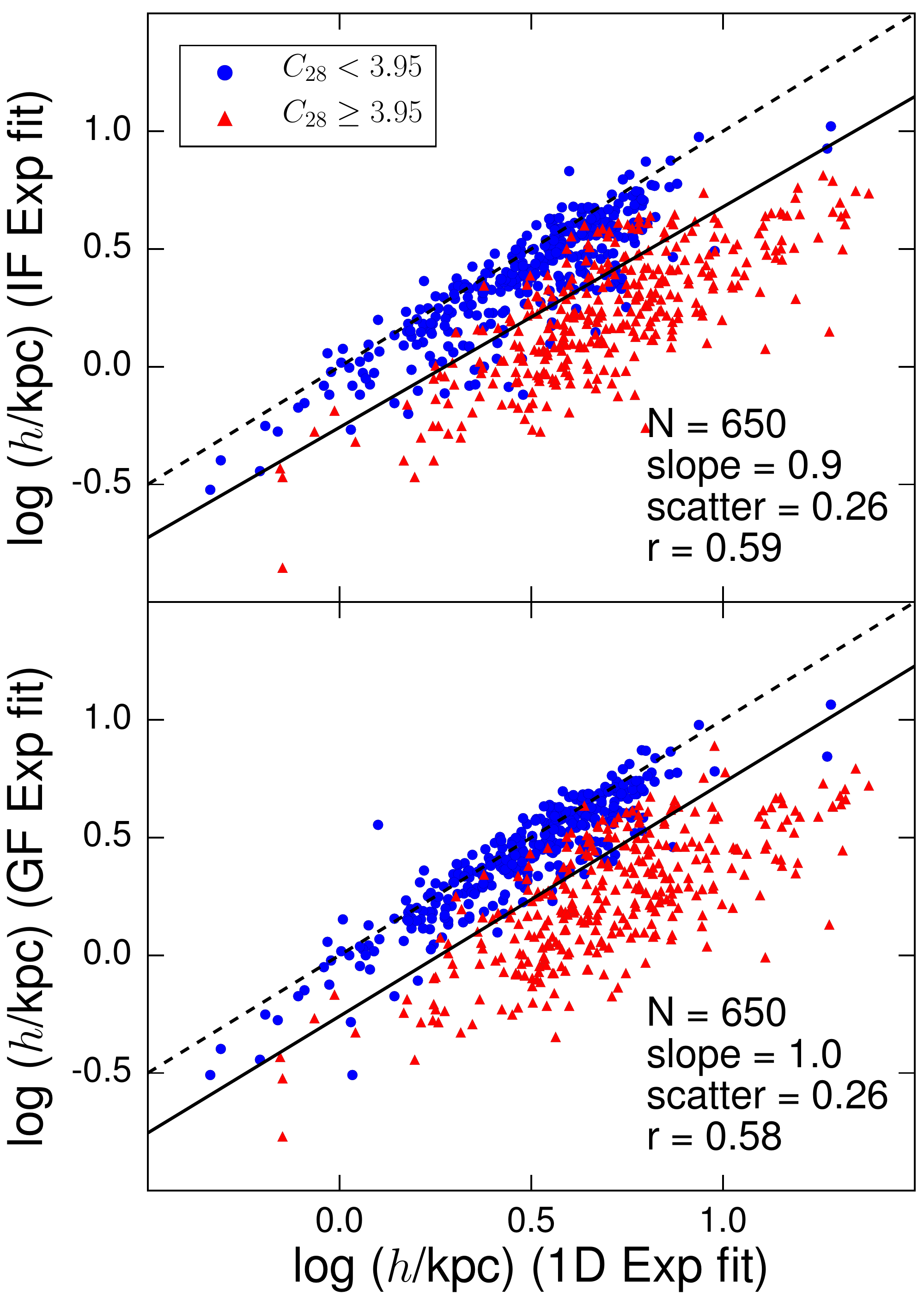}
\caption{Comparison of disc scale lengths of single Exponential models for all galaxies fitted in 2D
with \textsc{imfit} (IF) and \textsc{galfit} (GF), and in 1D.
The dashed lines indicate a 1:1 match while the solid lines show the best linear fit to all galaxies.}
\label{fig:hd_2D_vs_1D}
\end{centering}
\end{figure}

The assessment of the 1D and 2D models can be broken into three main components.
First the 1D and 2D models must be judged in a relative sense: Do the 1D and 2D 
instances of the same model produce similar parameters? Are the same models identified 
as preferred in 1D and 2D? 
After establishing any agreement between the 1D and 2D decompositions,
we examine the correlation strength of the resulting structural parameters with the
circular velocity metric V50$_{c}$ or velocity dispersion $\sigma$ -- both independent,
non-photometric parameters.  These will inform the preference for either 1D or 2D modelling
and will place their utility in context with the non-parametric quantities extracted from surface brightness
profiles (see \se{eval_1D_2D}). 
We also consider the bulge-disk ratio in \se{bd_ratios} to compare the relative stability and reliability of 
bulge modelling approaches (both in choice of bulge model and modelling software).

We first address the simplest model, a single Exponential function.  
The best fit scale lengths $h$ of 1D and 2D models for all galaxies are compared in \fig{hd_2D_vs_1D}. 
The low concentration systems display
reasonable agreement between 1D and 2D models. However, more highly concentrated systems
are significantly separated from the 1:1 line. Our 2D measurements
of $h$ obtained with \textsc{imfit} and \textsc{galfit} are comparable, with 0.11 dex of scatter.

This behaviour suggests different weighting between the inner and outer regions of galaxies. 
Because the 1D fits are carried out in magnitude units, the galaxy outskirts are (log-)weighted 
as strongly as in the brighter inner regions. Effectively, the 1D single Exponential fits
to high concentration systems attempt to fit an average slope that describes no part of the galaxy well.
In contrast, \textsc{imfit} and \textsc{galfit} ultimately favor the inner region at the expense the galaxy's outskirts for
high concentration systems. The bright central region of a galaxy is expected to have higher
individual pixel weights, and the greater area of the outskirts is expected to balance this effect. 
However, this concentration-dependent difference between 1D and 2D results suggests some 
overall bias towards the inner regions of galaxies. 

\begin{figure}
\begin{centering}
\includegraphics[width=0.49\textwidth]{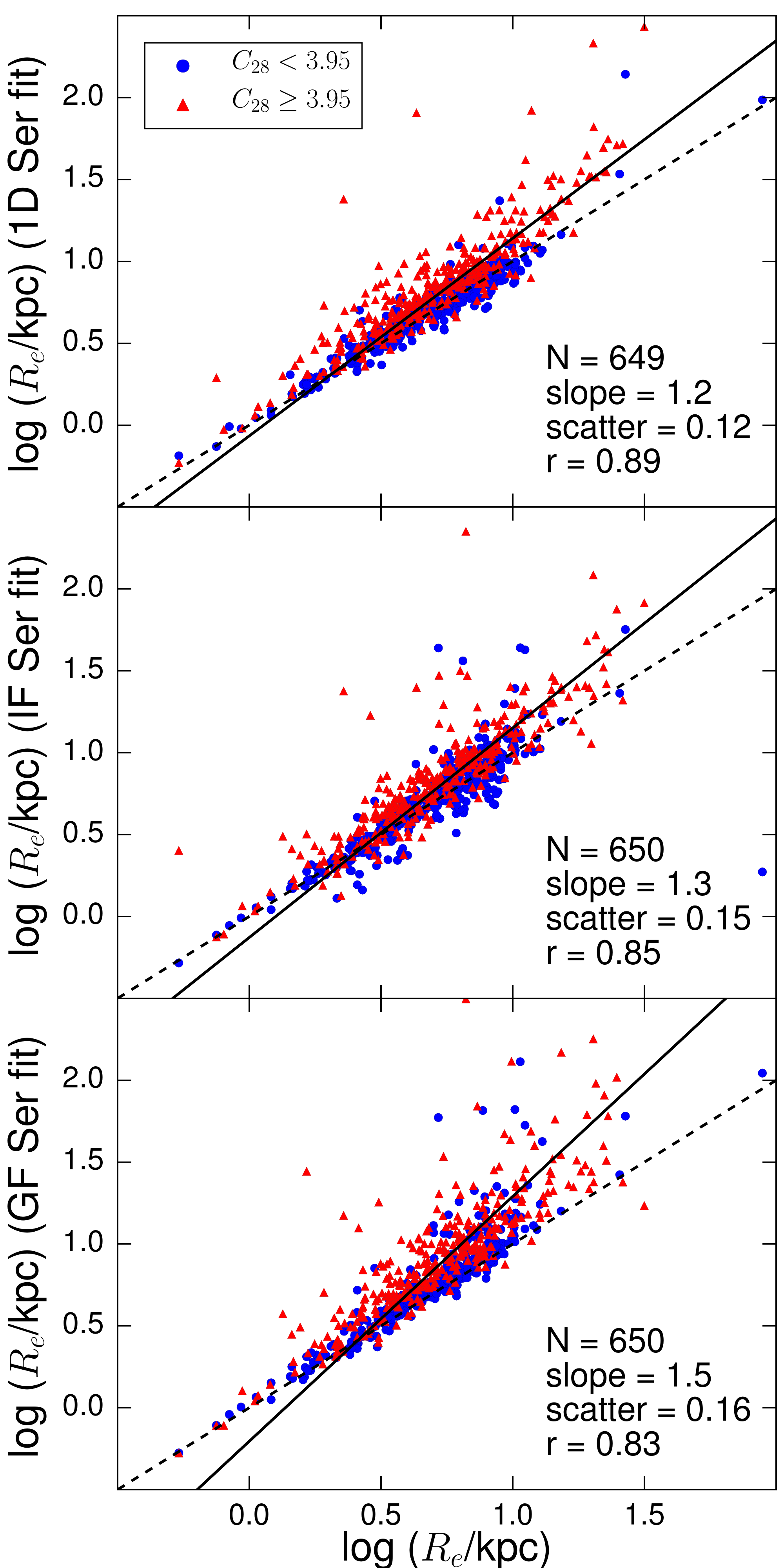}
\caption{Comparison of effective radii of single S\'ersic models for all galaxies fitted in 1D and in 2D
with non-parametric effective radii. The dashed lines indicate a 1:1 match while the solid lines
show the best linear fits.}
\label{fig:Re_2D_vs_1D}
\end{centering}
\end{figure}

These seemingly different weighting schemes do not imply that 2D models will
systematically misrepresent a galaxy's outskirts.
Indeed, the 1D, \textsc{imfit}, and \textsc{galfit} S\'ersic fitted effective radii $R_e$ are 
nearly all comparable to the non-parametric value, $R_e$, shown in  \fig{Re_2D_vs_1D}. 
There is a tendency towards larger modelled $R_e$ with higher concentration
for both 1D and 2D fitting results.  The 1D single S\'ersic fits provide the 
closest $R_e$ measurements to those obtained non-parametrically from
the surface brightness profiles. 

One must recall that these plots feature 1D and 2D fits for all galaxies, not necessarily
those that are well-described by these models. The continued discussion below and
in Appendix~\ref{sec:eval} address the correlations and properties of preferred models alongside
those of each model as applied to all galaxies in this sample.

\begin{figure*}
\begin{centering}
\includegraphics[width=0.32\textwidth]{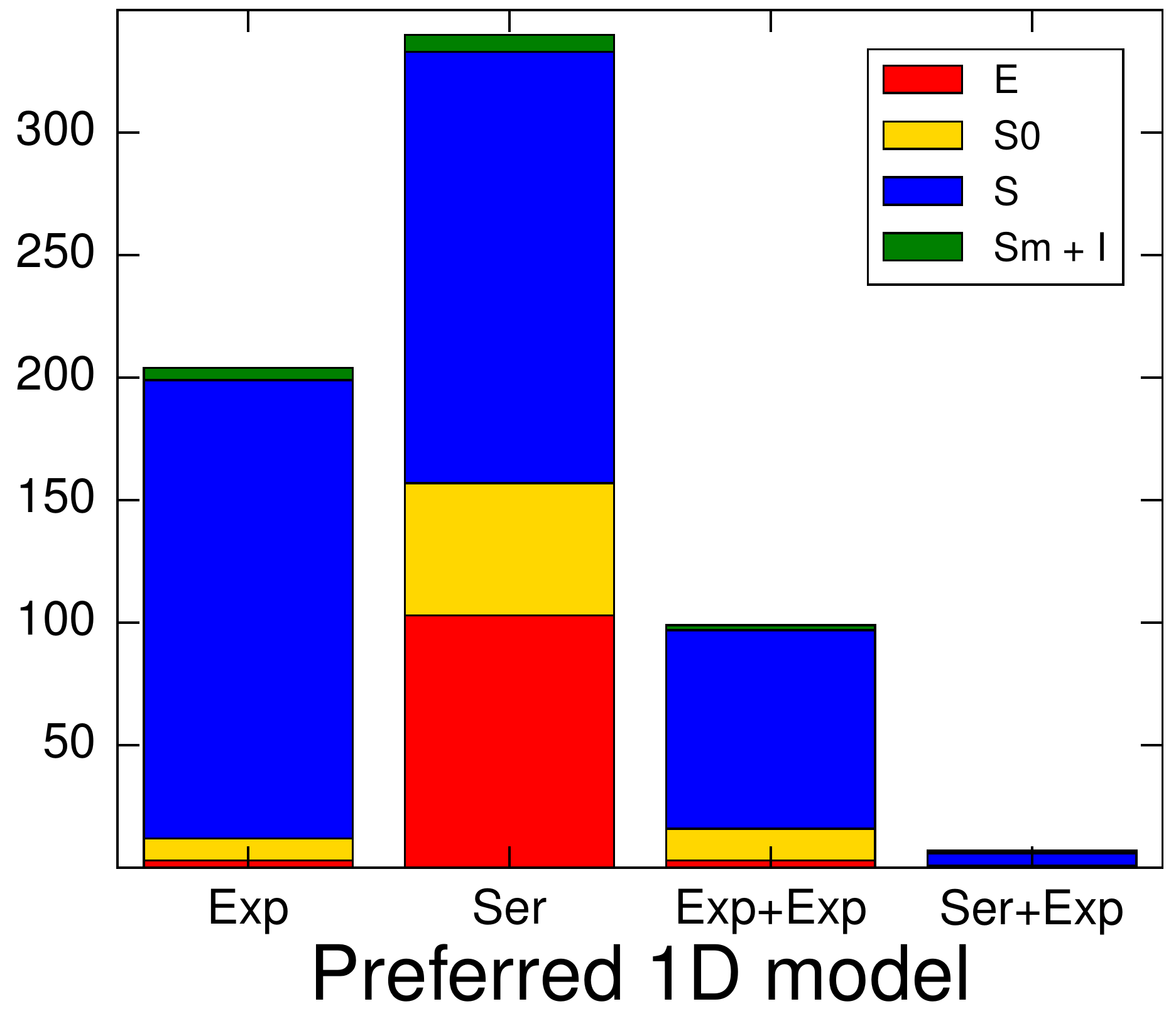}
\includegraphics[width=0.32\textwidth]{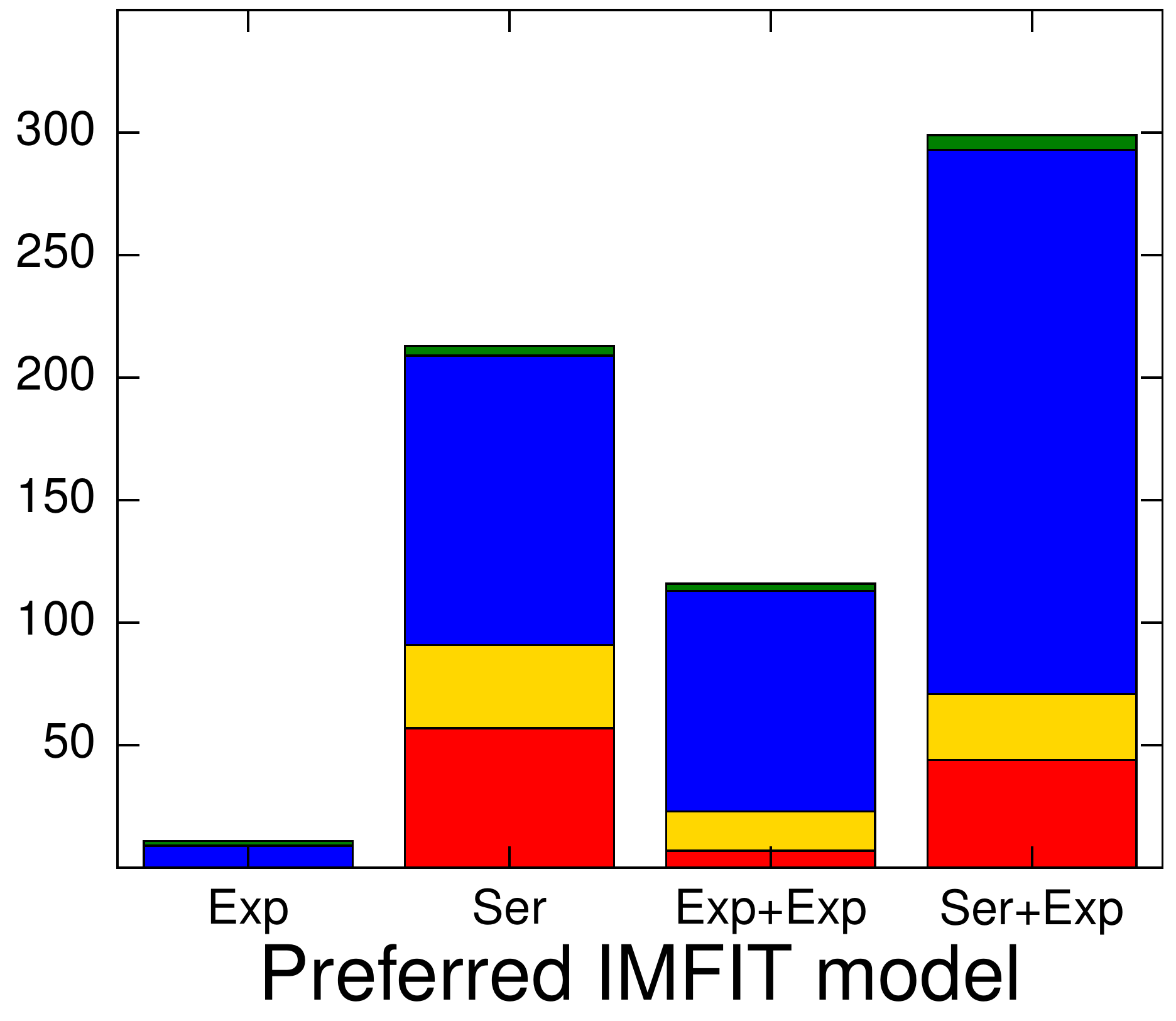}
\includegraphics[width=0.32\textwidth]{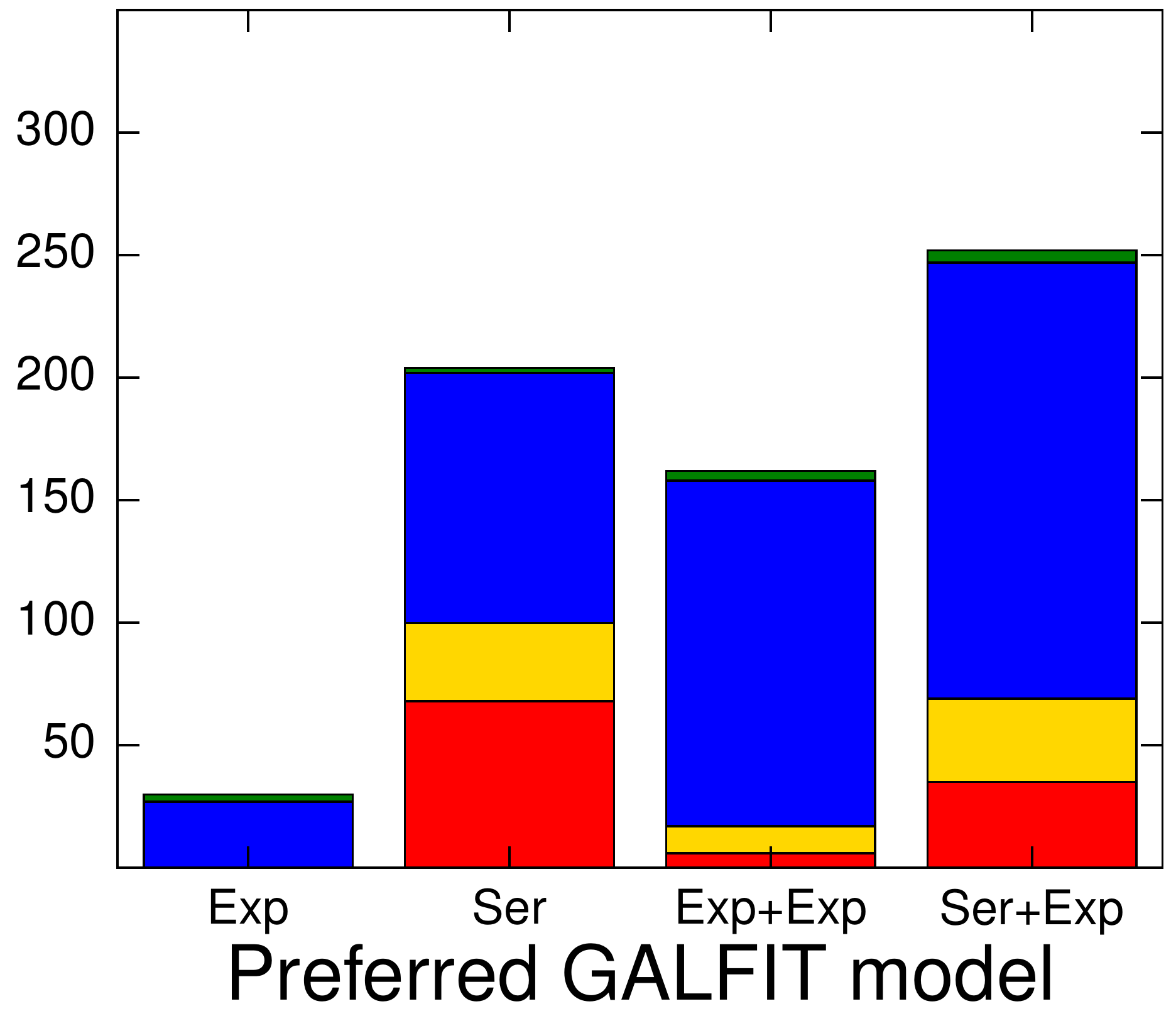}
\caption{Distribution of preferred 1D and 2D models.}
\label{fig:pref_models}
\end{centering}
\end{figure*}

The preferred models identified using 1D and 2D data do not agree well for most galaxies.
Only 167 out of 650 well-masked galaxies have the same preferred model when using 1D fits
and \textsc{imfit}, with another 129 preferring a single-component or two-component model
for both.  For \textsc{galfit}, 185 well-masked galaxies have the same preferred model as with
1D fits and another 131 are found to be best described by a 
single-component model or two-component model in both. The distribution
of preferred models in \fig{pref_models} illustrates the origin of this disagreement.
The 2D single component models (especially the Exponential model) are not as successful
at describing a galaxy and very rarely have a lower reduced $\chi^2$ than a two-component
model. The addition of a second component counteracts the tendency to favour the 
bright inner regions by enabling dimmer features to be fit alongside the brightest ones.
Additionally, the greater success of 2D two-component models is likely attributable
to their better ability to handle two regimes of position angle and ellipticity,
unlike single-component models which have one global position angle
and ellipticity (unless tilted rings are employed).

Among the 1D preferred models, elliptical galaxies mostly prefer a
S\'ersic model. This is expected, given their high concentration relative to spiral
galaxies and their curved surface brightness profiles. Many lenticular
galaxies share this 1D parametric classification. While the S\'ersic model
is the favoured \textsc{imfit} and \textsc{galfit} model for early-type galaxies,
this is by a fine margin. Roughly half of early-type galaxies are actually best
described by two-component 2D models. This may reflect the reduced
sensitivity at low surface brightnesses for \textsc{imfit} and \textsc{galfit} relative to our 1D fits;
it may be necessary to probe the variation in slope of a S\'ersic function at mid to
outer radii to correctly identify its global parameters. Otherwise, slight deviations
from the idealized distribution in the bright inner regions may lead the fit astray. 

The higher proportion of 2D preferred S\'ersic + Exponential models relative to double
Exponential models indicates that the flexible shape of the bulge component significantly
increases the goodness of fit. In contrast, the S\'ersic + Exponential model is the least frequent
preferred 1D model. This may be partly due to many 1D S\'ersic + Exponential
models abusing the S\'ersic component to fit overall curvature or the excess light of
a spiral arm rather than acting as a bulge component, earning the model a quality flag of 1
and excluding it from eligibility as a preferred model. On a related note, the large number
of spiral galaxies found to be best described by a 1D S\'ersic function is suspicious and
may be a reflection of the function's power to describe a variety of shapes and variations
beyond the classic de Vaucouleurs profile, potentially including what a human operator
would identify as a bulge and variations in the outer disc.  The ability to consider all radial
regimes of a galaxy equally may therefore lead to unexpected outcomes. The model with
the lowest reduced $\chi^2$ may not be the most morphologically suitable.

\begin{figure}
\begin{centering}
\includegraphics[width=0.45\textwidth]{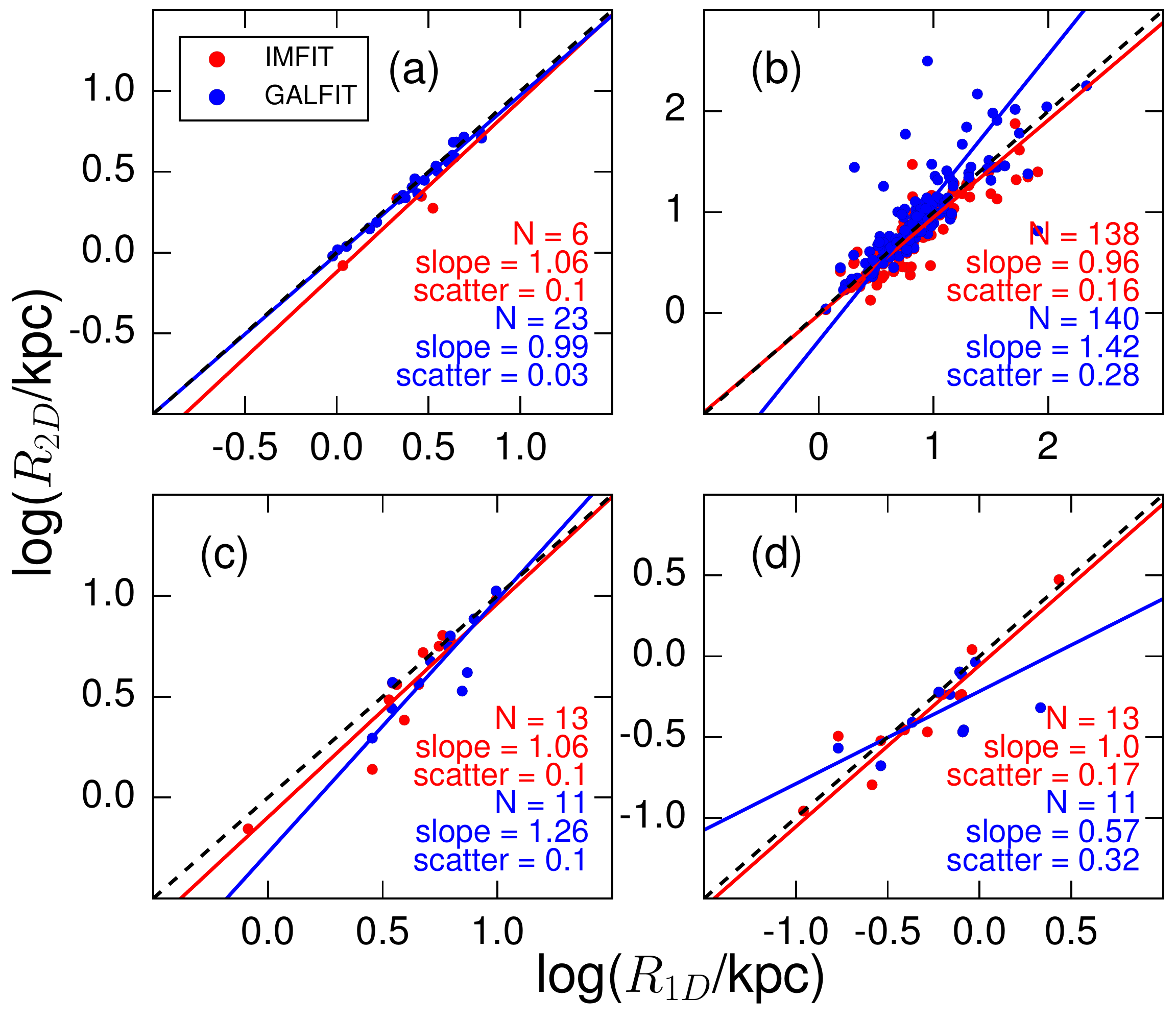}
\caption{Comparison of length parameters of various models for galaxies with the same prefered model in 1D and 2D (using \textsc{imfit}).
The dashed lines indicate a 1:1 match while the solid lines show the best linear fit through all the points. 
(a) Comparison of single Exponential scale lengths for galaxies best described by this model in both 1D and 2D.
(b) Comparison of single S\'ersic effective radii for galaxies best described by this model in both 1D and 2D.
(c) Comparison of Double Exponential disc scale lengths for galaxies best described by this model in both 1D and 2D. 
(d) Comparison of Double Exponential bulge scale lengths for galaxies best described by this model in both 1D and 2D. }
\label{fig:R_2D_vs_R_1D}
\end{centering}
\end{figure}

When the same preferred model is selected using both 1D and 2D methods (either \textsc{imfit} or \textsc{galfit}),
a fair agreement in the length parameters measured by both methods is found.  These comparisons can be seen
in \fig{R_2D_vs_R_1D}.  This is expected to hold when comparing 1D and 2D models with quality flags of 3.
The bulge scale length in the Double Exponential model is more poorly constrained than the
disc scale length, indicated by the increased scatter in \fig{R_2D_vs_R_1D}, panel (d) compared
to panel (c).  This is readily explained by the smaller influence of the bulge on global fit statistics,
leading to greater uncertainty in the bulge parameters. 

\subsection{IMFIT versus GALFIT}

Overall, our \textsc{imfit} and \textsc{galfit} models have been found to be largely equivalent,
relative to  our 1D models and non-parametric $R_e$. Qualitatively, \textsc{galfit} seems less impacted by
increasing concentration when attempting to describe low surface brightness features (reflected
in the greater number of successful Exponential models compared to \textsc{imfit}). 

Any difference between \textsc{imfit} and \textsc{galfit} must lie in the generation of sigma images,
given that the same images, masking, PSFs, gain, parameter bounds, and initial parameters were used. 
\textsc{imfit} makes use of the original sky level of the images while \textsc{galfit} does not; we are
unaware of other major differences. 
Although the adopted minimization techniques differ (Nelder-Mead for \textsc{imfit}
versus Levenberg-Marquardt for \textsc{galfit}), this is unlikely the source of any differences.

In practical terms, \textsc{galfit} may offer some advantages over \textsc{imfit}, such as:

\begin{itemize}
\item Input and output files share common formatting and contain all information required to run \textsc{galfit};
\item Faster execution times (under similar running conditions);
\item Surface brightness (central or effective) or luminosity can be chosen for model normalization;
\item Multiple options for parameter constraints are available;
\item Complex model components such as generalized ellipses, bending modes, and spirals are available.
\end{itemize}

Likewise, \textsc{imfit} distinguishes itself from \textsc{galfit} in a few ways:

\begin{itemize}
\item Object-oriented design facilitates easy creation of new model components;
\item Multiple options for minimization algorithms are available;
\item Multiple options for estimating parameter uncertainties (including Markov Chain Monte Carlo) are available;
\item Includes simple stand-alone executable for model image generation.
\end{itemize}

\subsection{Bulge-disc ratios}
\label{sec:bd_ratios}

\begin{figure}
\begin{centering}
\includegraphics[width=0.5\textwidth]{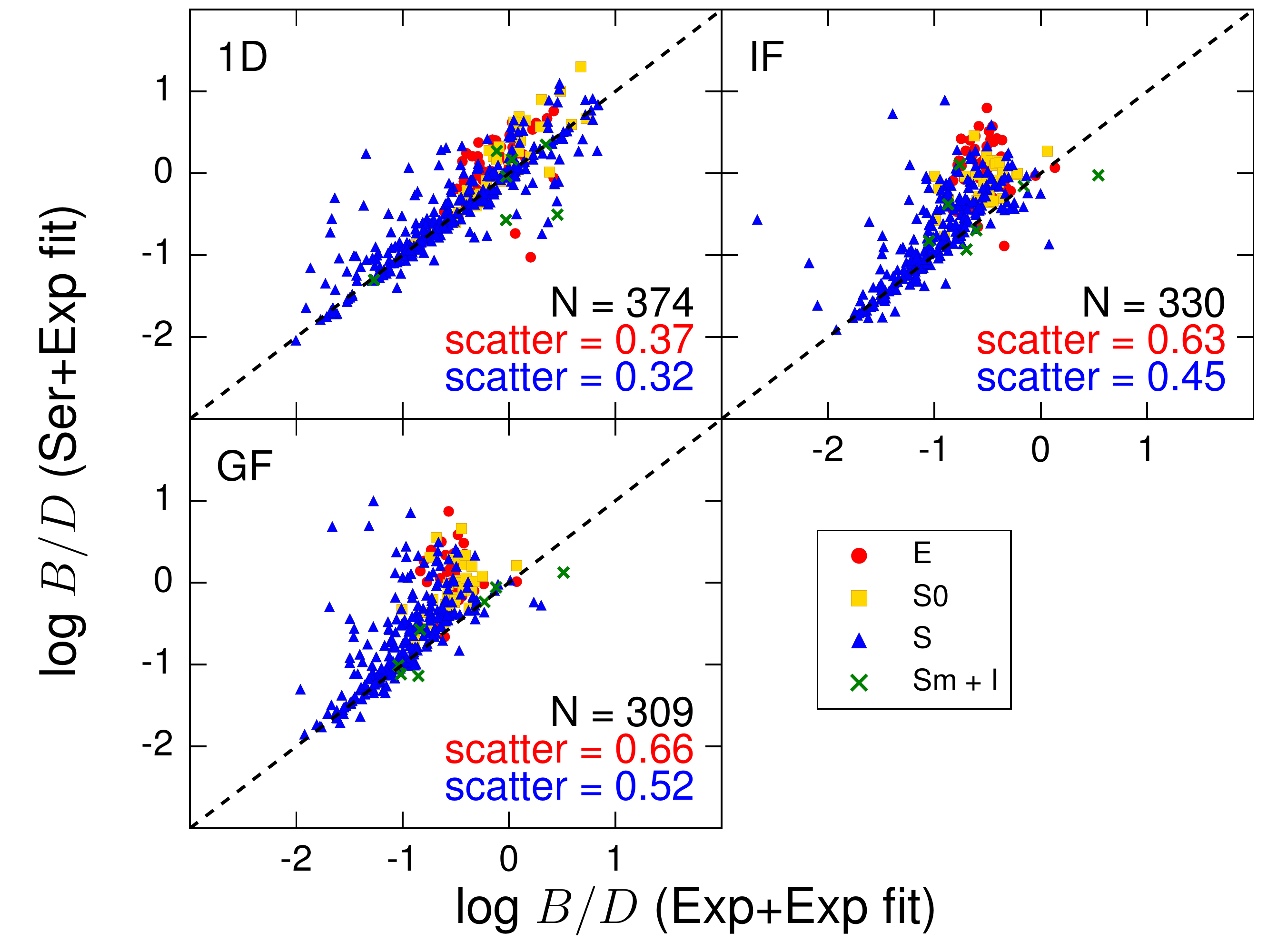}
\caption{Comparison of bulge-disc ratios with a S\'ersic or Exponential bulge model
for our 1D fits, as well as our \textsc{imfit} (IF) and \textsc{galfit} (GF) 2D models.
The disk model is exponential in all cases.
The scatter in each panel for early and late types about the
1:1 dashed line is shown in red and blue, respectively.}
\label{fig:bd_src_vs_exp}
\end{centering}
\end{figure}

The bulge-disc ratio ($B/D$ ratio) of a galaxy is the ratio of the flux contributed by its bulge and disc components.
This is most easily measured by fitting a bulge and disc model to the light profile or image of the galaxy. 
As a morphological indicator, $B/D$ ratios can vary from $>1$ for bulge-dominated systems to 0 for pure
disc systems. Measurements of $B/D$ ratios may depend heavily on the models adopted for the bulge and disc components, 
and will have different physical meaning when applied to early-type galaxies as they are highly concentrated
but well-described by a single smooth component. 

We examine $B/D$ ratios here to test of the reliability of the S\'ersic + Exponential model.
Since $n$ may be poorly constrained \citep{macarthur03}, leaving it as a free parameter may have impacted our results. 
\fig{bd_src_vs_exp} compares $B/D$ ratios when fitting S\'ersic and exponential
bulges, for 1D fits, \textsc{imfit}, and \textsc{galfit}.  Only galaxies with a significant bulge and significant disc component 
($0.001 < B/D < 10$) fitted and acceptable quality flags for both
bulge-disc models are considered.

For all modelling approaches, the adoption of a S\'ersic bulge over an Exponential bulge yields systematically
larger $B/D$ ratios. This effect is more dramatic for early-type galaxies, likely because the underlying light
distribution is S\'ersic-like and the switch to a S\'ersic bulge gives the latter component the flexibility to
match the light distribution over a larger range of radii. In 1D, early-type systems with a Double Exponential 
$B/D > 0.3$ ($\log B/D > -0.5$) form a loose sequence parallel to a 1:1 relation, with roughly 0.5 dex larger $B/D$ ratio
for S\'ersic bulges. For the lower $B/D$ ratios, there is fair agreement between the two bulge models.
For \textsc{imfit} and \textsc{galfit}, the early-type galaxies are scattered more widely and do not display
a region of better agreement towards low $B/D$ ratios. Many late-type galaxies cluster along the 1:1 line but
 the rest display high scatter. This clustering is limited to $B/D_{Exp} < 0.1$ for our 2D models but extends over a larger
range of $B/D$ ratios for our 1D models. Overall, the agreement of the 2D $B/D$ ratios for our two bulge models seems to be limited by 
the lack of large $B/D$ ratios among the Double Exponential models.

In many cases, exponential and S\'ersic bulges are not interchangeable. On average, the adoption
of a S\'ersic bulge model yields a brighter bulge over the exponential case. This is especially true for
early-type galaxies; however, these are often best described by a single S\'ersic model and should
not be modelled with a bulge-disc model without strong motivation. 
We call for caution in the choice of model components and the interpretation of $B/D$ ratios.

\subsection{Advantages and disadvantages of 1D and 2D modelling}

Whether one prefers 1D or 2D image modelling is somewhat subjective: the 1D approach requires
the extraction of azimuthally-averaged surface brightness profiles though 1D profile modelling is straightforward
and quick. Likewise, 2D image modelling is easily implemented, but the increased execution time is
significant compared to 1D modelling without providing new physically meaningful insights
unless highly detailed models are developed under close supervision.  1D
surface profiles also offer additional information about radially changing ellipticities and position angles. 

2D modelling does offer greater ease at modelling additional non-axisymmetric components
such as bars, rings, and spiral arms that might not deproject along circular isophotes.  1D
azimuthally-averaged isophotal solutions, especially those with fixed position angle and ellipticity,
will typically smear out those features.  Conversely, we have found the simplest single
component models to be poorly executed with 2D methods relative to 1D modelling.

The parameters of the 2D models fitted here are either consistent with their 1D model
counterparts or are more weakly correlated with an unbiased metric where the two sets
of models disagree (see Appendix~\ref{sec:eval_1D_2D}).  This would thus favor a 1D approach, whenever possible. 

If one wishes to describe the global properties of a galaxy, it is desirable to give uniform
consideration to the inner and outer regions of a galaxy.  This must therefore be taken
into consideration when considering 1D versus 2D modelling. Some of the improved sensitivity
to low surface brightness features with 1D modelling may be attributed to the profile extraction
method rather than the explicitly uniform weighting at all radii.
The use of a fixed position angle and ellipticity during profile extraction in the outermost
regions of a galaxy has been shown to improve profile depth \citep{courteau96}.
No equivalent treatment is given to the galaxy outskirts in 2D modelling to tease
the faint galaxy signal out of the sky.


\section{Comparison with existing CALIFA studies}
\label{sec:compare}

\subsection{ Effective radii}
\label{sec:compare_Re}

\begin{figure}
\begin{centering}
\includegraphics[width=0.46\textwidth]{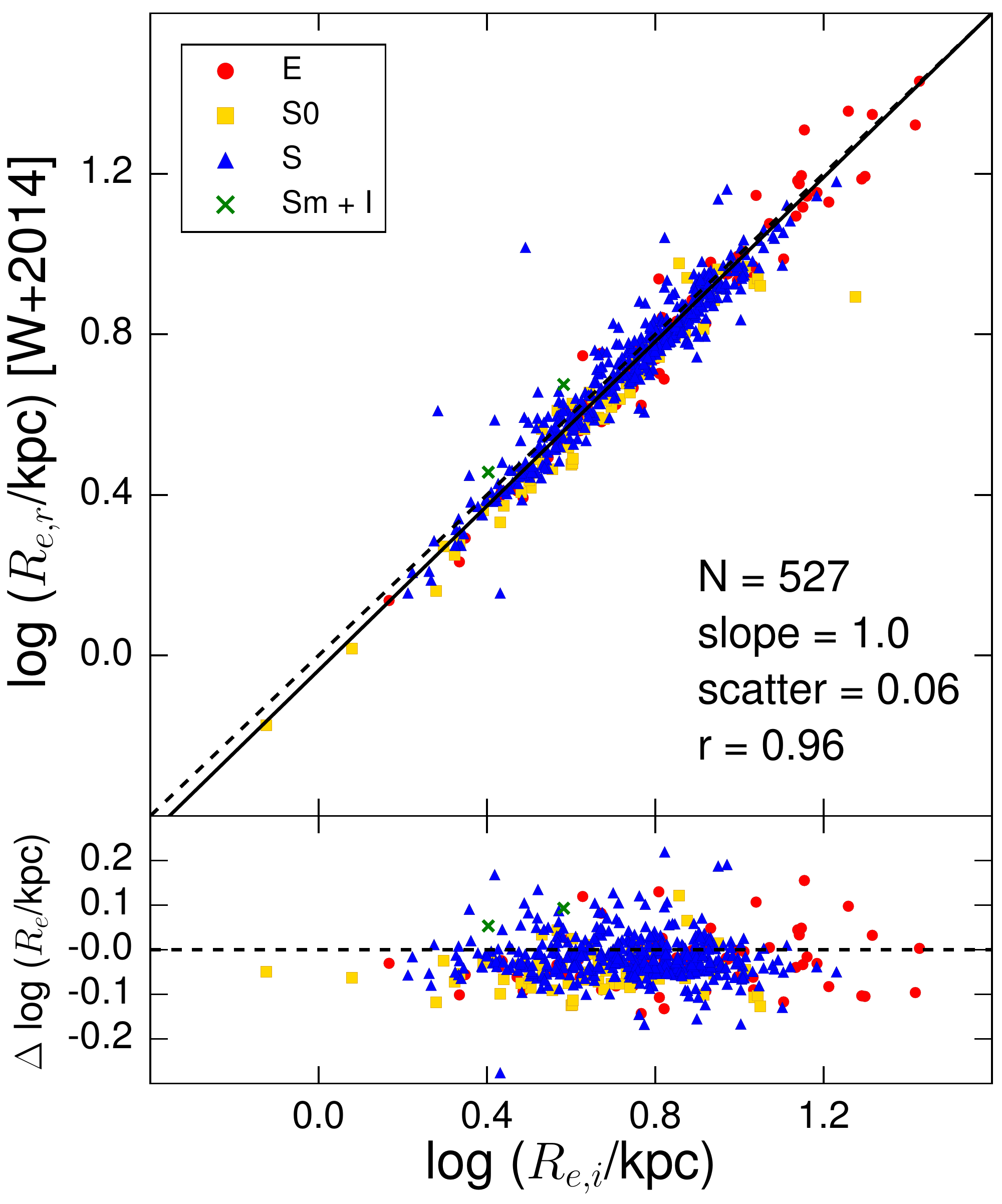}
\caption{Comparison of effective radii extracted by \protect\cite{walcher14} (vertical axis) and ourselves (horizontal axis). 
The bandpasses for each of the effective radius measurements are labeled with the subscripts $r$ and $i$.
The dashed line indicates a 1:1 match while the solid line shows the best linear fit through all the points. 
Residuals are shown in the bottom panel.}
\label{fig:Re_W14_vs_CG}
\end{centering}
\end{figure}

Our measured $R_e$ can be compared to those calculated by \cite{walcher14}. 
The latter also make use of SDSS imaging and fit isophotal contours to obtain a light curve of growth.
Our approaches differ mainly on account of their constant position angle and ellipticity for the
entire galaxy and the measurement of $R_e$ in the $r$-band instead of the $i$-band.  Furthermore, Walcher
et al. do not tabulate isophotal radii or model the light profiles.  \fig{Re_W14_vs_CG} shows a very close match
between our respective $R_e$, with a fairly low scatter of 0.06 dex. This suggests that any potential bias
introduced when smoothing isophotes or selecting best contours by eye is marginal, at least as applied
following our approach.

\begin{figure}
\begin{centering}
\includegraphics[width=0.46\textwidth]{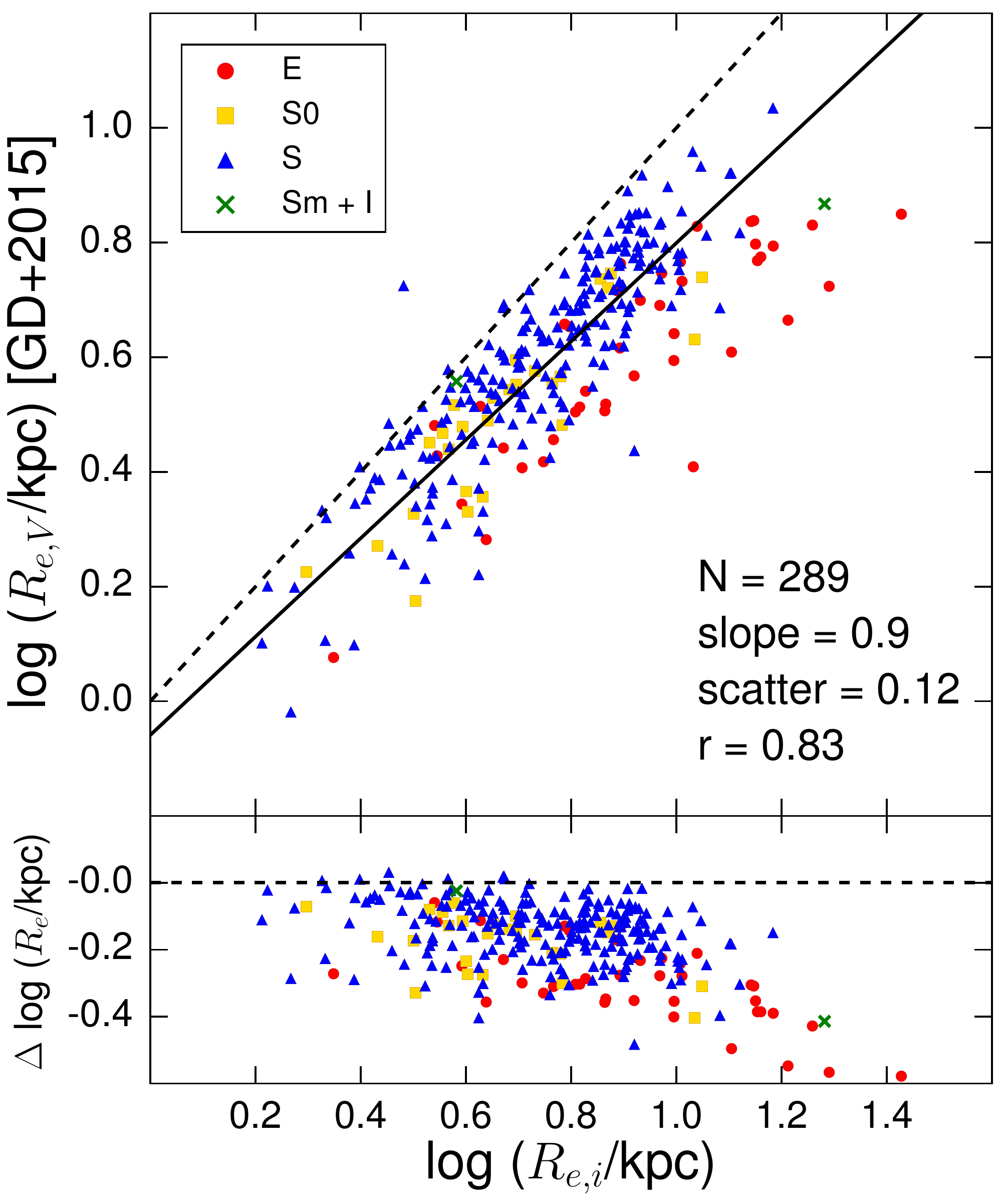}
\caption{Comparison of effective radii extracted from CALIFA datacubes \citep[vertical axis]{gonzalezdelgado15} and ourselves
(horizontal axis). The bandpasses for each of the effective radius measurements are labeled with the subscripts $V$ and $i$.
The dashed line indicates a 1:1 relation while the solid line shows the best linear fit through all the points.
Residuals are shown in the bottom panel.}
\label{fig:Re_GD15_vs_CG}
\end{centering}
\end{figure}

\cite{gonzalezdelgado15} calculated $R_e$ using CALIFA data by searching for the
elliptical aperture that contains half of the total light at 5635 \AA{}.  This definition clearly differs from
ours and \cite{walcher14} and consistently underestimates $R_e$, as shown in \fig{Re_GD15_vs_CG}. 
The limited field of view of CALIFA compared to SDSS imaging can explain
the offset.  While galaxies in the CALIFA survey are size-selected to sample the PPAK IFU most efficiently,
the outskirts of the galaxy can fall outside of its field of view and the total amount of light is therefore
underestimated, yielding a smaller estimate of $R_e$.  This discrepancy naturally increases for larger galaxies,
explaining why the best-fitted slope in \fig{Re_GD15_vs_CG} is less than unity. The differing
bandpass may also contribute to the differences between our $R_e$.

Early-type galaxies seem to define a separate, shallower sequence than the later types, indicating that
their effective radii are more greatly underestimated. Early types have typically higher S\'ersic $n$,
thus resulting in a greater underestimation of the total light when using CALIFA IFU data, and therefore
a greater underestimation of $R_e$, than for late-type galaxies with $n\sim1$. 

\subsection{Stellar masses}
\label{sec:compare_Mstar}

\begin{figure}
\begin{centering}
\includegraphics[width=0.46\textwidth]{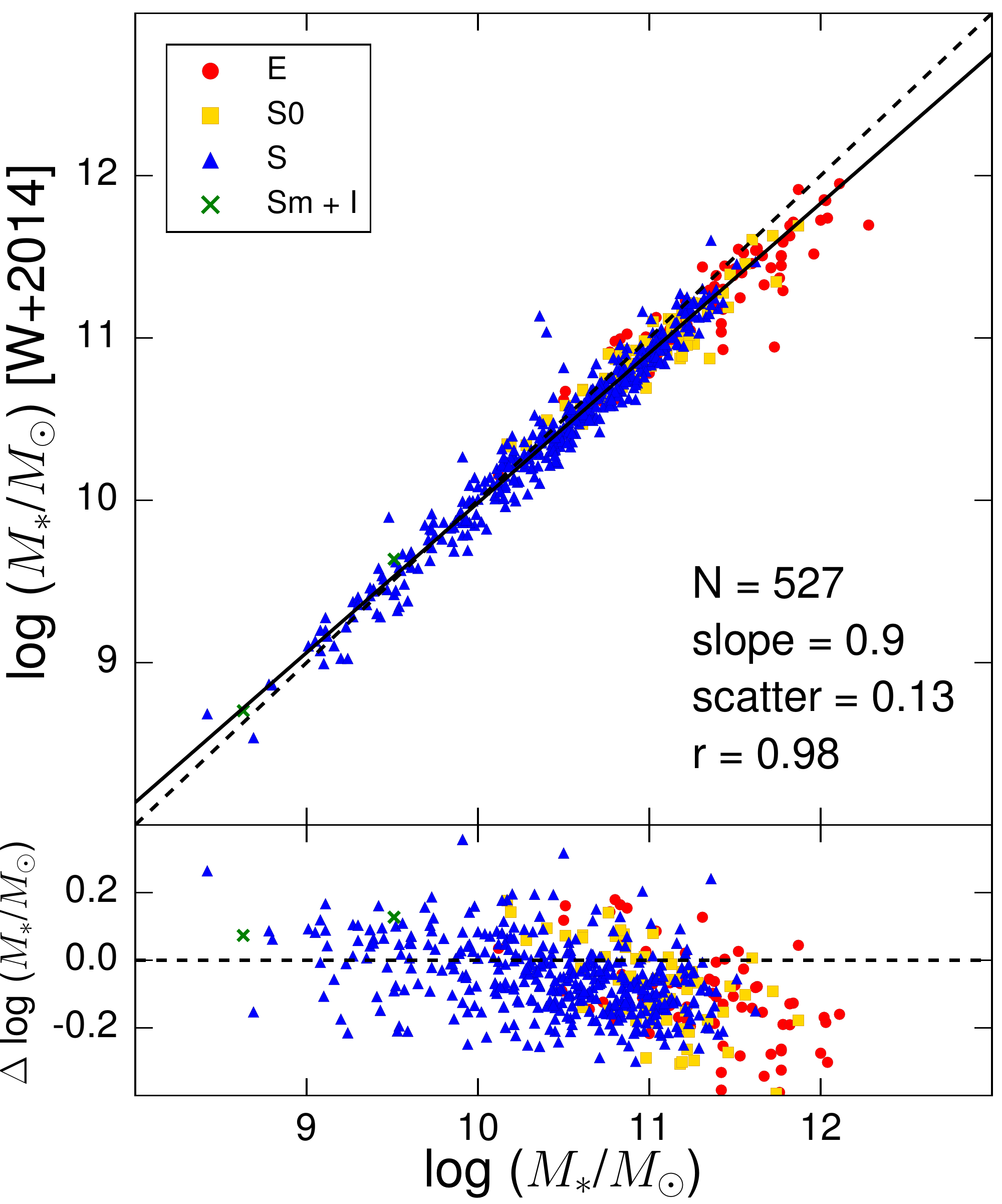}
\caption{Comparison of our stellar masses (horizontal axis) with those calculated by \protect\cite{walcher14} (vertical axis).
The dashed line indicates a 1:1 match while the solid line shows the best linear fit through all the points.
Residuals are shown in the bottom panel.}
\label{fig:Mstar_W14_vs_CG}
\end{centering}
\end{figure}

\begin{figure}
\begin{centering}
\includegraphics[width=0.46\textwidth]{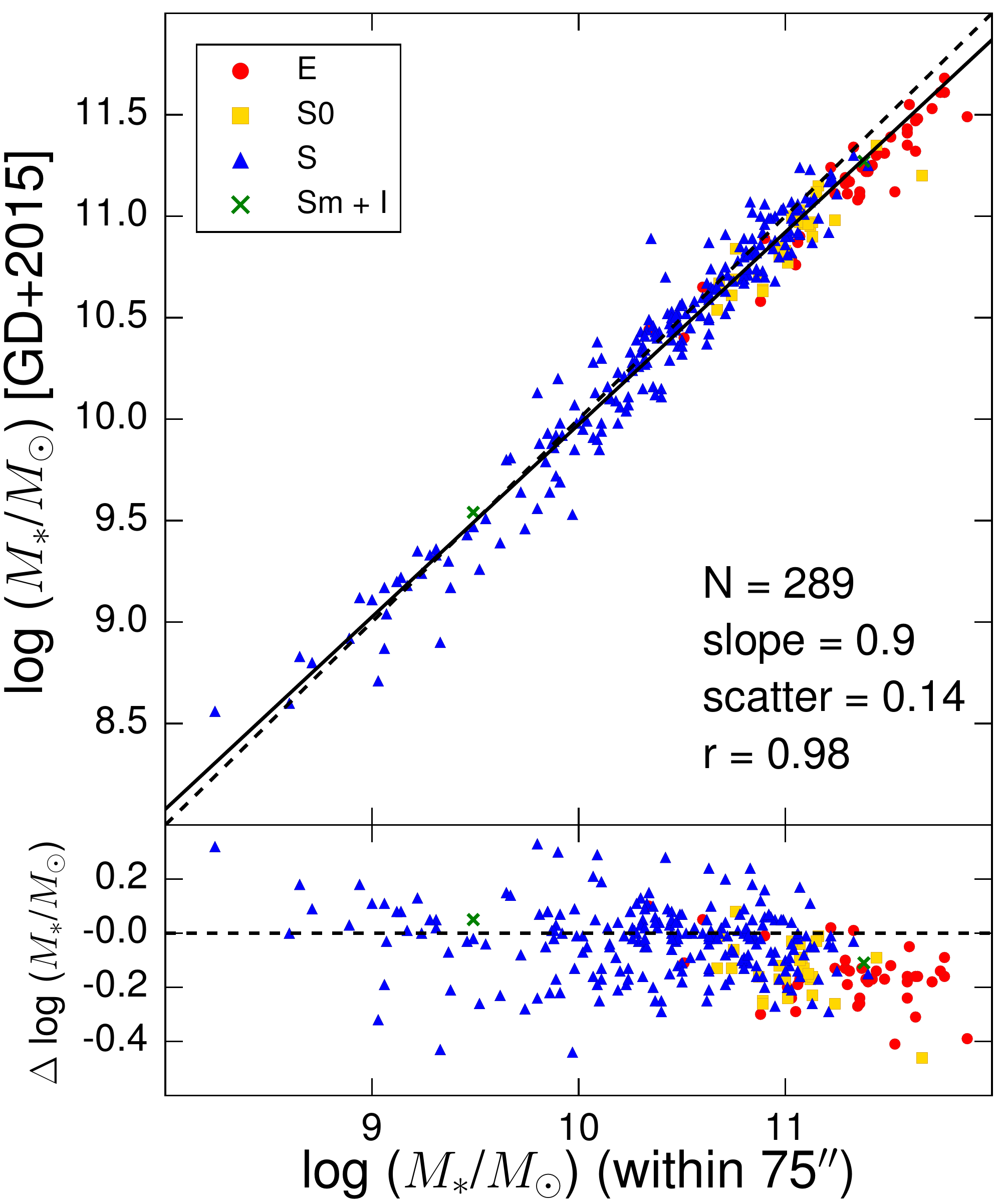}
\caption{Comparison of our stellar masses using SDSS $gri$ photometry with a limited aperture
 (horizontal axis) with those based on CALIFA data \citep[vertical axis]{gonzalezdelgado15}.
The dashed line indicates a 1:1 match while the solid line shows the best linear fit through all the points.
Residuals are shown in the bottom panel.}
\label{fig:Mstar_GD15_aperture}
\end{centering}
\end{figure}%

\begin{figure}
\begin{centering}
\includegraphics[width=0.46\textwidth]{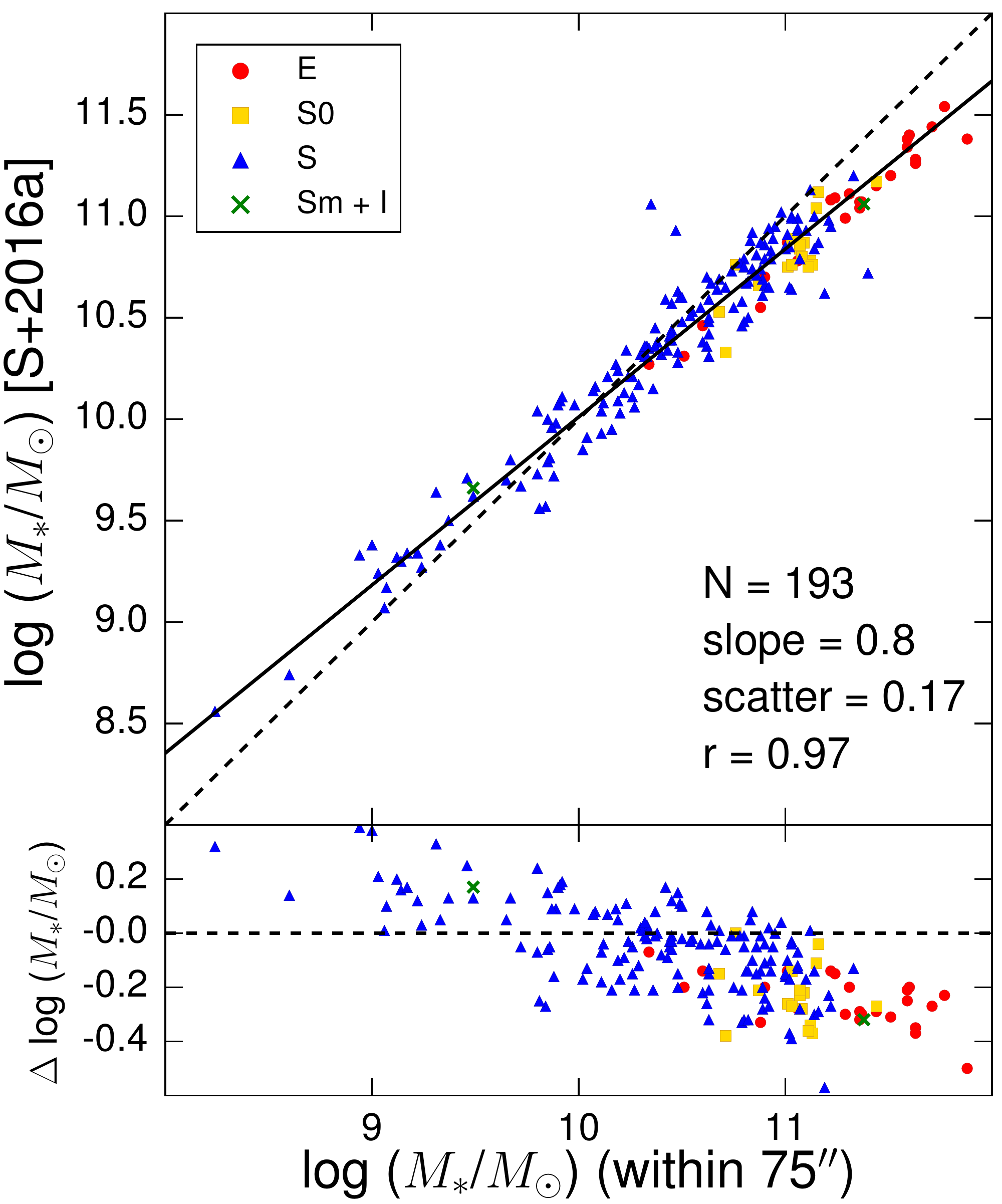}
\caption{Comparison of our stellar masses using SDSS $gri$ photometry with a limited aperture
(horizontal axis) with those based on CALIFA data \citep[vertical axis]{sanchez16}. 
The dashed line indicates a 1:1 match while the solid line shows the best linear fit through
all the points. Residuals are shown in the bottom panel.}
\label{fig:Mstar_S16_aperture}
\end{centering}
\end{figure}

To further validate our photometric catalog and investigate the impact of CALIFA's
limited aperture size, we now compare our stellar masses with those in existing
databases. First, we consider the stellar masses tabulated by \cite{walcher14}.
In addition to a matching
SDSS photometric source and surface brightness extraction methods, we use the same
stellar population models \citep{bruzual03} to extract stellar masses. We both
calculate stellar masses using the total integrated light of our surface
brightness profiles. Whereas Walcher et al.
use spectral energy distribution (SED) fitting to all five SDSS bands, we employ 
MLCRs instead. Despite these slightly differing methods, a close match with
a scatter of 0.13 dex is found for our stellar masses (see \fig{Mstar_W14_vs_CG}).
This successful comparison further bolsters the merits of our photometric catalog.

Our stellar masses may also be compared with those obtained from spatially resolved
stellar population modelling of CALIFA data.  
We define a simple ansatz for aperture matching to investigate the possibility of aperture effects
on stellar masses. We limit our surface brightness profiles to 37.5$\arcsec$ in radius,
or 75$\arcsec$ in diameter, corresponding to the largest dimension of the CALIFA PPak IFU field of view.
The proposed comparison is not a strictly rigorous match, since the CALIFA field of view is hexagonal.
The inclination and position angle of disc galaxies thus also play a role in filling up the available area.
The fact that galaxies are more concentrated in mass than in light \citep{gonzalezdelgado14}
will minimize the impact of the restricted CALIFA aperture on measurements of stellar mass,
compared to the dramatic changes in $R_e$ seen above.

\cite{gonzalezdelgado15} use two SSP libraries in their work. The CBe library uses updated 
\cite{bruzual03} SSPs and a Chabrier IMF; results from this library yield greatest similarity 
with the MLCRs of \cite{roediger15}. We find a good match between our stellar masses and those of 
\cite{gonzalezdelgado15}, with poorer agreement at low and high stellar masses. The trend to 
lower CALIFA-based masses at high stellar mass suggests an aperture-related bias. At the low mass 
end, galaxies have mostly lower CALIFA-based masses. This difference could be related to the 
greater depth of SDSS imaging compared to CALIFA rather than an aperture bias. 
When we recalculate our stellar masses within a limited aperture, the overall scatter
is reduced by 0.02 dex and a closer 1:1 match is achieved (though the best-fitting linear slope remains
unchanged). The comparison of our limited aperture stellar masses to those of \cite{gonzalezdelgado15}
is shown in \fig{Mstar_GD15_aperture}. From this exercise, we expect that stellar masses determined 
from either stellar population modelling of optical spectra or SED fitting to broadband optical 
imaging are almost equivalent \emph{when aperture and stellar population models are the same.} 

A similar comparison with stellar masses obtained by \cite{sanchez16} yields a tilted relation,
with the best agreement at intermediate masses. The match is not as tight as with \cite{gonzalezdelgado15}
due to our differing stellar population models. The above speculation on the source of the mass discrepancies holds here as well.
Adopting an aperture-limited stellar mass for this comparison, shown in  \fig{Mstar_S16_aperture},
results in a 0.01 dex decrease in scatter and slightly closer to 1:1 agreement for massive galaxies. 
However, the relation's slope has not changed and remains shallower than unity.  This indicates that the 
differences in stellar population models are a more significant source of disagreement than differences in aperture.

While some differences exist, the overall agreement between stellar masses obtained from
broadband optical photometry MLCRs, SED fits, or full optical spectra is rather satisfactory.
Given that stellar spectra are less readily available, it is indeed comforting that photometric 
assessment yield reliable stellar masses. The impact of aperture size on CALIFA-derived
structural measurements highlights the importance of wide-field broadband photometry to
support ongoing and future IFS surveys.

\subsection{Comparison with \protect\cite{mendezabreu17}}
\label{sec:compare_models}

Our 1D and 2D parametric models can be compared with the 2D models of \cite{mendezabreu17},
fitted using GASP2D \cite{mendezabreu08}. 
These authors provide a best model per galaxy, fitted in $g$, $r$, and $i$ SDSS images. 
Given the great range in the choice of model components, a few simplifications are made to identify reasonable matches.
We neglect nuclear bars and nuclear point sources, and consider a disc + bar (+ bulge) model to be equivalent to a
S\'ersic + Exponential model. In the cases where both single S\'ersic and a bulge-disc model are offered in the
catalog, the model with lowest $i$-band $\chi^2$ is taken to be the preferred one.  There are no true matches for our 
Double Exponential model as \cite{mendezabreu17} do not consider an exponential bulge model; rather, we allow partial 
matches with their bulge-disc models. 

Our catalogs and final models are quite different, especially in light of the different suites of model components.
For the 404 galaxies modelled by \cite{mendezabreu17}, 128 match our preferred 1D models, 217 match our preferred
\textsc{imfit} models, and 223 match our preferred \textsc{galfit} models. 
The low number of matches with our preferred 1D models is certainly due to the large number of galaxies best fitted
by a single Exponential or S\'ersic function when using that technique (\fig{pref_models}). It is expected that human
supervision and a wider suite of model components will lead to the adoption of more complicated models. 
Our preferred models are more similar when considering \textsc{imfit} and \textsc{galfit}, though again the
observed propensity towards selecting simple single S\'ersic models limits the number of matching preferred
models between our works. 

We can identify a closely matching model from our models catalog for 171 out of 404 galaxies modelled by
M\'endez-Abreu et~al., with another 105 having approximate matches outlined by our criteria above. For these
matching models, there was generally strong agreement between length and brightness parameters,
and weaker agreement between S\'ersic $n$ when applicable.  
st{For the conversion of relevant parameters
to absolute sizes using redshift and plotted against V50$_{c}$, the correlation strengths for our respective
measured parameters are roughly equivalent, suggesting that one set of models is not strongly favoured
over the other.} 
We note that the disc models of M\'endez-Abreu et~al. are somewhat dimmer, as expected by 
considering additional components: inner bars and nuclear point sources.  

It is worth noting that while M\'endez-Abreu et al. extract 1D surface brightness profiles in order to initialize their 2D fits, 
a catalog of 1D model parameters is not presented.  Likewise, only their preferred model is made available to the reader.


\section{Summary}
\label{sec:discuss}

We have presented a diverse and homogeneous photometric catalog of 667 CALIFA galaxies,
covering the entirety of the third and final data release.  A variety of non-parametric structural
parameters were extracted from azimuthally averaged surface brightness profiles.  Where overlap
between \cite{walcher14} and \cite{mendezabreu17} exists, a generally good agreement is found,
especially with the former. Our photometric and modelling catalog is more extensive than any
existing resource for CALIFA DR3 galaxies, especially when combined with our $gri$ surface
brightness profiles.  Our catalog is made available to the community in order to provide a reliable,
comprehensive, and transparent photometric reference to complement and support any
exploitation of the spectroscopic CALIFA data base.

A great majority of CALIFA galaxies are adequately described by a single Exponential or a single
S\'ersic model when modelling their 1D surface brightness profiles, but these single-component
models are not as successful when modelling galaxy images directly due to an apparent
tendency to fit the bright inner regions at the expense of the rest of the galaxy. 
We recommend the use of multi-component models for any 2D galaxy modelling
exercises in order to mitigate this behaviour. The addition of a nuclear point source may potentially
improve this situation, though we have not made use of such a component here.
For detailed multi-component models of CALIFA galaxies, see \cite{mendezabreu17}.
Investigating multi-component models revealed that disc parameters are better constrained than
bulge parameters, and that the bulge-disc ratio is highly sensitve to the choice of a bulge model. 

\subsection{The importance of large spatial coverage}

The bias in $R_e$ calculated from spatially-limited CALIFA data \citep{gonzalezdelgado15} reinforces
the crucial supporting role of broadband photometry with generous spatial coverage. 
The dynamics
derived from CALIFA may also be limited, particularly for the largest spiral galaxies on the sky.
Their rotation curves may continue to rise beyond CALIFA's coverage, resulting in the underestimation
of $V_{max}$. 

Agreement between our computed stellar masses and those of \cite{walcher14} is quite good;
however, a less satisfactory match was found with various CALIFA-derived stellar masses \citep{gonzalezdelgado15,sanchez16}.
Recalculating our stellar masses over matching radial ranges probed by CALIFA resulted in improved agreement.
Aperture-induced biases in CALIFA-derived stellar masses are thus significant despite earlier claims 
\citep{gonzalezdelgado15}. However, we note that the choice of stellar population
models is much more influential in the agreement or disagreement of various stellar mass measurements. 

\section*{Acknowledgments}

\noindent
CG and SC acknowledge support from the Natural Science and Engineering Research Council (NSERC)
of Canada through a PGS D scholarship and a Research Discovery Grant, respectively.  This
work also benefited from enlightening discussions with 
Peter Erwin, Jairo M\'endez-Abreu, Sebastian F. S\'anchez, and Jakob Walcher.
Our referee's comments contributed to a leaner presentation of this paper.

Funding for the Sloan Digital Sky Survey IV has been provided by
the Alfred P. Sloan Foundation, the U.S. Department of Energy Office of
Science, and the Participating Institutions. SDSS-IV acknowledges
support and resources from the Center for High-Performance Computing at
the University of Utah. The SDSS web site is www.sdss.org.

SDSS-IV is managed by the Astrophysical Research Consortium for the 
Participating Institutions of the SDSS Collaboration including the 
Brazilian Participation Group, the Carnegie Institution for Science, 
Carnegie Mellon University, the Chilean Participation Group, the French Participation Group, Harvard-Smithsonian Center for Astrophysics, 
Instituto de Astrof\'isica de Canarias, The Johns Hopkins University, 
Kavli Institute for the Physics and Mathematics of the Universe (IPMU) / 
University of Tokyo, Lawrence Berkeley National Laboratory, 
Leibniz Institut f\"ur Astrophysik Potsdam (AIP),  
Max-Planck-Institut f\"ur Astronomie (MPIA Heidelberg), 
Max-Planck-Institut f\"ur Astrophysik (MPA Garching), 
Max-Planck-Institut f\"ur Extraterrestrische Physik (MPE), 
National Astronomical Observatory of China, New Mexico State University, 
New York University, University of Notre Dame, 
Observat\'ario Nacional / MCTI, The Ohio State University, 
Pennsylvania State University, Shanghai Astronomical Observatory, 
United Kingdom Participation Group,
Universidad Nacional Aut\'onoma de M\'exico, University of Arizona, 
University of Colorado Boulder, University of Oxford, University of Portsmouth, 
University of Utah, University of Virginia, University of Washington, University of Wisconsin, 
Vanderbilt University, and Yale University.

This work has made use of the NASA/IPAC Extragalactic Database (NED) 
which is operated by the Jet Propulsion Laboratory, California Institute of Technology, under contract with 
the National Aeronautics and Space Administration.



\bibliographystyle{mnras}
\bibliography{CALIFA_photometry}



\appendix

\section{Correlations of photometric parameters with line widths}
\label{sec:eval}

Our study is largely motivated by the need for a robust and homogeneous catalog
of structural parameters for all CALIFA galaxies with SDSS DR10 imaging.  In the interest
of identifying the quantities that lead to the tightest scaling relations, we also test
their correlation with an independent, non-photometric parameter known to describe
the structure of all galaxies.
Since the potential, or mass, of a galaxy is known to play a fundamental role in the evolution
of these systems, the independent spectroscopically-determined
parameter of choice for gas-rich or gas-poor systems is the circular velocity or stellar
velocity dispersion, respectively. 

For gas-rich systems, we use the catalogue of \hi~ line widths by \cite{springob05}. 
Specifically, we choose W50, the width of the \hi~ line at 50 per cent of the peak flux,
divided by two, to obtain an estimate of the peak rotational velocity of the galaxy in
the plane of the sky.  The line widths are corrected for instrumental effects, redshift 
broadening, and turbulence; we further correct for projection using our own
photometric inclinations. Our halved and deprojected W50 are noted as V50$_{c}$.
Line widths are available for 253 out of the sample of 650 well-masked CALIFA galaxies.
Of these galaxies, 195 have inclinations in the range $40^\circ < i < 80^\circ$ and thus
present the most reliable V50 measurements and deprojections.

While this subsample has poor coverage of early-type galaxies, we still examine
the correlation of our photometric parameters with CALIFA velocity dispersions produced
by Pipe3D \citep[][Gilhuly et al., in prep]{sanchez16}.   We select the velocity dispersion
measured in absorption that is typically the most reliable for early-type galaxies: that
derived with highest spectral resolution ($R\sim 1650$, corresponding to the V1200 grating)
and measured within a 30$''$ aperture.  We note this dispersion as $\sigma_{30}$.

\subsection{Radii, magnitudes, and stellar masses}
\label{sec:eval_R}

\begin{table}
\begin{centering}
\begin{tabular}{c c c c c}
\hline \hline
Parameter & $N$ & slope & scatter (dex) & $r$ \\ \hline \hline
$R_e$  & 190 & 1.5 & 0.15 & 0.46 \\
$R_{23.5}$  & 190 & 1.3 & 0.12 & 0.7 \\
\\
$M_{23.5}$  & 190 & -9.5 & 0.1 & 0.8 \\
$M_i$  & 190 & -9.2 & 0.1 & 0.8 \\
$\log(M_*/M_\odot)$  & 190 & 4.5 & 0.1 & 0.79 \\
\hline \hline
\end{tabular}
\caption{Correlations of length parameters from 1D and 2D models with V50$_c$.
These correlations are computed using all galaxies for which a reliable deprojected V50$_{c}$ is available. }
\label{tab:phot_corr}
\end{centering}
\end{table}

\begin{table}
\begin{centering}
\begin{tabular}{c c c c c}
\hline \hline
Parameter & $N$ & slope & scatter (dex) & $r$ \\ \hline \hline
$R_e$  & 131 & 2.2 & 0.15 & 0.57 \\
$R_{23.5}$  & 131 & 1.2 & 0.14 & 0.69 \\
\\
$M_{23.5}$  & 131 & -7.4 & 0.13 & 0.71 \\
$M_i$  & 131 & -7.8 & 0.13 & 0.71 \\
$\log(M_*/M_\odot)$  & 131 & 3.3 & 0.12 & 0.76 \\
\hline \hline
\end{tabular}
\caption{Correlations of length parameters from 1D and 2D models with $\sigma_{30}$.
These correlations are computed for all early-type galaxies for which $\sigma_{30}$ is available. }
\label{tab:phot_corr_sigma}
\end{centering}
\end{table}

Tables~\ref{tab:phot_corr} and \ref{tab:phot_corr_sigma} summarize the correlation
strengths of several key structural parameters with V50$_c$ and $\sigma_{30}$, respectively.
The correlations seen here are generally strong; indeed, these are reflections of the
size-velocity and mass-velocity (or Tully-Fisher) relations. 
 
Of the two characteristic radii extracted, $R_{23.5}$ correlates most strongly
with HI line width. This likely indicates that $R_{23.5}$ is more closely related
to galaxy mass or luminosity and is therefore a better descriptor of the global
properties of the galaxy.  $R_e$ displays a somewhat weak correlation with V50$_{c}$
due to its dependence on apparent morphology alone. Two galaxies with differing
total luminosity may have the same $R_e$ if the shape of their light profiles is identical
(albeit with a different normalization). The reduced sensitivity with galaxy mass or luminosity
introduces scatter to the correlation.  On the other hand, this makes $R_e$ a useful physical
scale on which to compare galaxies of diverse sizes and morphologies. 

The correlation between our non-parametric radii with $\sigma_{30}$ reveals that $R_{23.5}$
still displays the tightest match. The correlation with $R_e$ is also stronger for the
earlier than the later types.

We find correlations of equal strength between each of $M_{23.5}$, $M_i$, and $M_*$ with
V50$_c$. As most of a galaxy's light comes from its inner brightest regions, it is anticipated that 
$M_{23.5} \sim M_{tot}$ and these magnitudes should yield equally strong correlations with V50$_{c}$.
Direct comparison of $M_{23.5}$ and $M_{tot}$ reveals an extremely tight
relation with a slope near unity. 

The corresponding exercise for early types shows weaker correlations. 
While mass and rotation velocity are sufficient to describe spiral galaxies, early-type galaxies require size
in addition to mass and velocity dispersion to reproduce their more representative Fundamental Plane 
\citep{djorgovski87,dressler87,bernardi03,cappellari06}.
Interestingly, $M_*$ correlates somewhat more strongly with $\sigma_{30}$ than either of the $i$-band magnitudes,
a distinction that is not seen with the late-type galaxies in \Table{phot_corr}. 

\subsection{Parametric models}
\label{sec:eval_1D_2D}

\begin{table}
\begin{centering}
\begin{tabular}{c c c c c c}
\hline \hline
Model & param & $N$ & slope & scatter (dex) & $r$ \\ \hline \hline
\multirow{2}{*}{1D Exp}  & \multirow{2}{*}{$h$} & 189 & 1.3 & 0.14 & 0.57 \\
                         &                        & 77  & 1.4 & 0.10 & 0.73 \\
\\
\multirow{2}{*}{IF Exp}  & \multirow{2}{*}{$h$} & 182 & 2.1 & 0.16 & 0.22  \\
                         &                        & 6 & -- & -- & -- \\
\\
\multirow{2}{*}{GF Exp}  & \multirow{2}{*}{$h$} & 190 & 2.8 & 0.16 & 0.26 \\
                         &                        & 10 & -- & -- & --  \\
\\
\multirow{2}{*}{1D Ser}  & \multirow{2}{*}{$R_e$} & 190 & 1.5 & 0.15 & 0.51 \\
                         &                        & 78  & 1.1 & 0.19 & 0.43  \\
\\
\multirow{2}{*}{IF Ser}  & \multirow{2}{*}{$R_e$} & 188 & 2.0 & 0.14 & 0.56 \\
                         &                        & 46 & 1.5 & 0.15 & 0.54  \\
\\
\multirow{2}{*}{GF Ser}  & \multirow{2}{*}{$R_e$} & 183 & 1.9 & 0.13 & 0.6  \\
                         &                        & 36 & 1.7 & 0.11 & 0.74  \\
\\
\multirow{2}{*}{1D Exp+Exp}  & \multirow{2}{*}{$h_d$} & 183 & 1.4 & 0.15 & 0.53 \\
                         &                            & 32  & 3.0 & 0.10 & 0.42 \\
\\
\multirow{2}{*}{IF Exp+Exp}  & \multirow{2}{*}{$h_d$} & 175 & 1.5 & 0.15 & 0.48  \\
                         &                            & 39 & 0.9 & 0.19 & 0.34  \\
\\
\multirow{2}{*}{GF Exp+Exp}  & \multirow{2}{*}{$h_d$} & 170 & 1.7 & 0.15 & 0.44 \\
                         &                            & 67 & 1.3 & 0.16 & 0.5  \\
\\
\multirow{2}{*}{1D Ser+Exp}  & \multirow{2}{*}{$h_d$} & 179 & 1.5 & 0.14 & 0.57  \\
                         &                            & 0  & -- & -- & -- \\
\\
\multirow{2}{*}{IF Ser+Exp}  & \multirow{2}{*}{$h_d$} & 157 & 1.6 & 0.14 & 0.57  \\
                         &                            & 93 & 1.4 & 0.12 & 0.63 \\
\\
\multirow{2}{*}{GF Ser+Exp}  & \multirow{2}{*}{$h_d$} & 143 & 1.5 & 0.13 & 0.55  \\
                         &                            & 76 & 1.8 & 0.12 & 0.51 \\
\hline \hline
\end{tabular}
\caption{Correlations of length parameters from 1D and 2D models with V50$_{c}$. These correlations
are computed for all galaxies (for which a reliable deprojected V50$_{c}$ is available) and then again
only for those galaxies that prefer the selected model.  Missing values occur for small samples
that lack meaningful measurements of slope, scatter, or Pearson $r$.}
\label{tab:model_W50_corr}
\end{centering}
\end{table}

\begin{table}
\begin{centering}
\begin{tabular}{c c c c c c}
\hline \hline
Model & param & $N$ & slope & scatter (dex) & $r$ \\ \hline \hline
1D Exp  & \multirow{3}{*}{$h$} & 129 & 2.2 & 0.11 & 0.69 \\
IF Exp  &   & 129 & 2.7 & 0.13 & 0.5  \\
GF Exp  &   & 130 & 2.4 & 0.13 & 0.52 \\
\\
1D Ser  & \multirow{3}{*}{$R_e$} & 129 & 4.0 & 0.13 & 0.53 \\
IF Ser  &   & 130 & 3.9 & 0.13 & 0.48 \\
GF Ser  &   & 130 & 4.5 & 0.13 & 0.49\\
\\
1D Exp+Exp & \multirow{3}{*}{$h_d$} & 130 & 2.4 & 0.12 & 0.66 \\
IF Exp+Exp &   & 127 & 2.4 & 0.13 & 0.56  \\
GF Exp+Exp &   & 124 & 2.4 & 0.13 & 0.57  \\
\\
1D Ser+Exp & \multirow{3}{*}{$h_d$} & 125 & 4.3 & 0.13 & 0.54\\
IF Ser+Exp &   & 108 & 4.3 & 0.13 & 0.52\\
GF Ser+Exp &   & 88 & 3.0 & 0.12 & 0.56  \\
\hline \hline
\end{tabular}
\caption{Correlations of length parameters from 1D and 2D models with $\sigma_{30}$. These correlations
are computed for all early-type galaxies (for which $\sigma_{30}$ is available). Since the majority
of these galaxies prefer the single S\'ersic model, no separate treatment for other models is
presented here as was done in \protect\Table{model_W50_corr}.}
\label{tab:model_sigma_corr}
\end{centering}
\end{table}

In addition to studying the correlation strength of non-parametric photometric parameters with
various characteristic velocities, we present the same treatment of our 1D and 2D model parameters.
These correlations are summarized in \Table{model_W50_corr} and \Table{model_sigma_corr}.

The single Exponential scale length, $h$, fitted in 1D correlates more strongly with V50$_c$
than that fitted with \textsc{imfit} or \textsc{galfit}.  These correlations strengthen when only preferred-model galaxies
are examined; the correlation with the 1D fitted $h$ is slightly stronger than that seen with $R_{23.5}$ ($r=0.70$). 

All fitted $R_e$ correlate more strongly with V50$_c$ than the non-parametric $R_e$ 
determined from the curve of growth. However, the non-parametric $R_e$
generally correlates more strongly with $\sigma_{30}$ than any fitted $R_e$. 
It is not clear whether fitted or non-parametric $R_e$ best describe a galaxy. Regardless,
$R_{23.5}$ and $h$ correlate more strongly with velocities for both early and late types
and are thus better characteristic sizes to use in the construction of tight scaling relations.

A fair correlation is seen between the disc scale length $h_d$ and V50$_c$
for the S\'ersic + Exponential model for 1D, \textsc{imfit}, and \textsc{galfit} 
models when all galaxies are considered. 
These correlation strengths are approximately equal to those seen for the single Exponential $h$. 
The disc scale length of the Double Exponential model shows similar though weaker correlation
strengths with V50$_c$. However, the correlation strength between $\sigma_{30}$ and
the Double Exponential disc scale length is stronger. 
In general, equal or greater correlation strength is obtained with the
single-component models over the disc scale lengths of the multi-component models. This
suggests that there is value in using simple models when parametrizing the global properties
of a galaxy within a statistical sample rather than attempting to fit more complex prescriptions.

Considering the correlation strength for scaling relations involving sizes for the whole sample,
 $R_{23.5}$ remains the radius of choice for all descriptions of galaxy structure.
The effective radius, $R_e$, is commonly used to normalize galaxies onto a common length scale. It may be worthwhile to 
consider $R_{23.5}$ for this purpose as well.

If only galaxies with a 1D single Exponential preferred model 
are studied, whose light profiles are surely close to pure exponential discs, their scale length $h$ 
is the most descriptive length tracer. This may benefit studies of scaling relations 
for well-behaved disc galaxies. Our 1D fitted $h$
also correlates strongly with $\sigma_{30}$ for early-type galaxies; indeed, stronger than any fitted or
non-parametric $R_e$. Users can refer to the model quality flags for the single Exponential model 
to verify whether the model traces only the bright inner regions of a galaxy or most/all of its visual extent.

\section{\texorpdfstring{\textit{\MakeLowercase{gri} }}~Surface brightness profiles} 
\label{sec:gri_profiles}

Final $gri$ surface brightness profiles for nine well-masked galaxies are plotted in \fig{gri}.
All profiles can be found online as supplementary material and at 
\url{https://www.physics.queensu.ca/Astro/people/Stephane_Courteau/gilhuly2017/index.html}.

\clearpage
\onecolumn 

\begin{figure}
\begin{centering}
\makebox[\textwidth][c]{\includegraphics[width=0.9\textwidth]{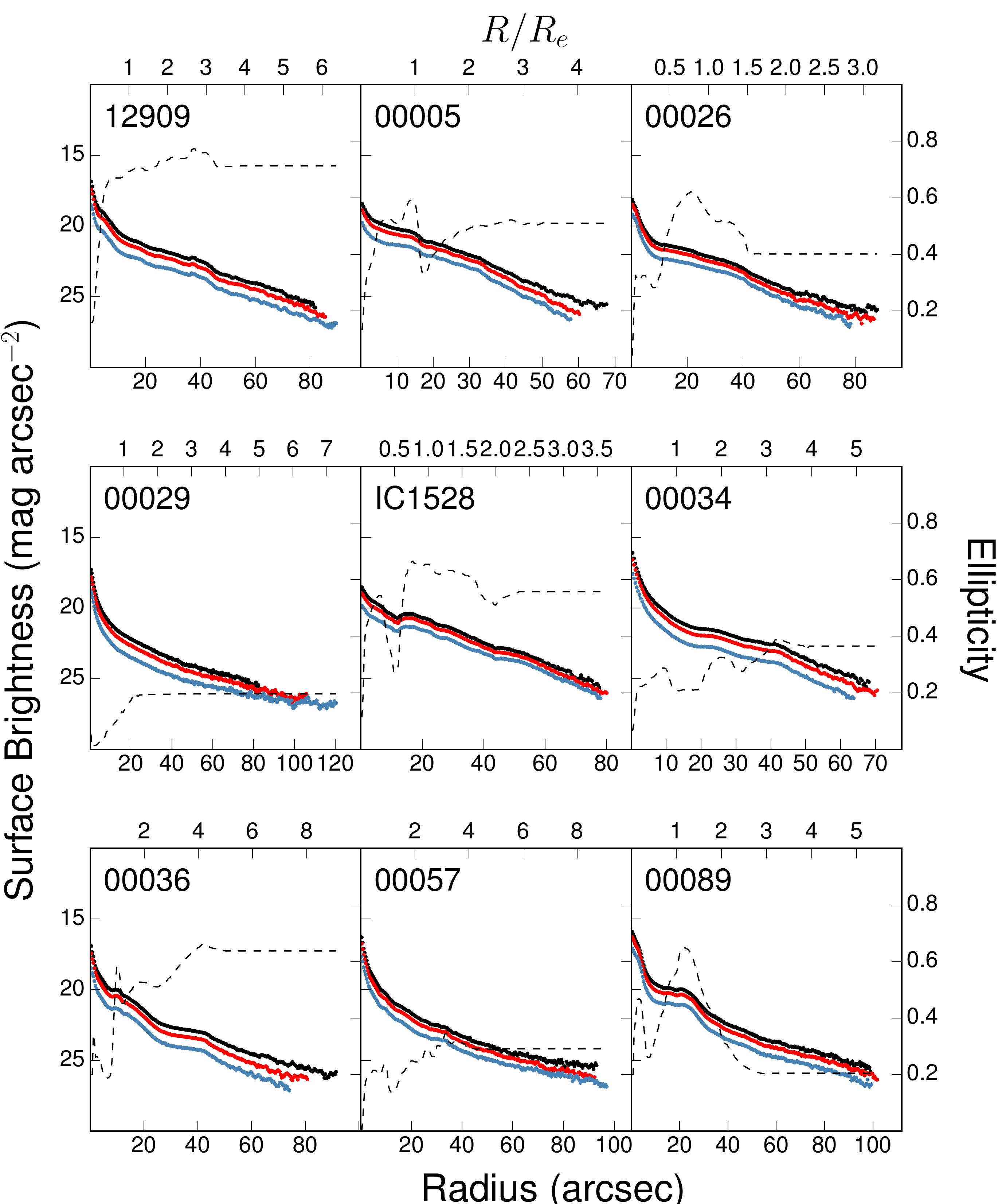}}
\caption{Blue, red, and brown points indicate the $g$-, $r$-, and $i$-band, respectively. 
Ellipticity profiles are shown with dashed black lines.  UGC numbers are found in the
top left corner of each panel; for galaxies with no UGC number, an alternate catalog
number is given. }
\label{fig:gri}
\end{centering}
\end{figure}


\bsp	
\label{lastpage}
\end{document}